\documentclass[prb,floats,twocolumn,aps]{revtex4-1}

\usepackage{graphicx}
\usepackage{amsfonts}
\usepackage{amssymb}
\usepackage{amsbsy}
\usepackage{amsmath}\usepackage[colorlinks = true,
            linkcolor = blue,
            urlcolor  = blue,
            citecolor = blue,
            anchorcolor = blue]{hyperref}
\usepackage{bm}

\begin{document}
\title{Interplay between spatial anisotropy and further
exchange interactions in the triangular Heisenberg model}

\author{M. G. Gonzalez, E. A. Ghioldi, C. J. Gazza, L. O. Manuel, and A. E. Trumper}

\affiliation {Instituto de F{\'i}sica Rosario (CONICET) and Universidad Nacional de Rosario, Boulevard 27 de Febrero 210 bis, (2000) Rosario, Argentina} 

\date{\today}

\begin{abstract} 
We investigate the interplay between spatial anisotropy and further exchange interactions in the spin-$\frac{1}{2}$ Heisenberg antiferromagnetic model on a triangular lattice. We use the Schwinger boson theory by including Gaussian fluctuations above the mean-field approach. The phase diagram exhibits a strong reduction of the long range collinear and incommensurate spirals regions with respect to the mean-field ones. This reduction is accompanied by the emergence of its short range order counterparts, leaving an ample room for $0$-flux and nematic spin liquid regions. Remarkably, within the neighborhood of the spatially isotropic line, there is a range where the spirals are so fragile that only the commensurate $120^{\circ}$ N\'eel ones survive. The good agreement with recent variational Monte Carlo predictions gives support to the rich phase diagram induced by spatial anisotropy.  
\end{abstract}

\maketitle 

\section{Introduction} 

Two dimensional (2D) frustrated magnets have been the natural playground for the search of non-conventional magnetic states like quantum spin liquids.\cite{Sachdev2008,Normand2009,Savary16,Zhou2017,Broholm2020}These states of matter are characterized by both, a strong quantum entanglement among the spins of different sites and the presence of fractional magnetic excitations. Unlike the magnetically ordered states, a quantum spin liquid state is topologically ordered in the sense that certain patterns of entangled spins emerge.\cite{Wen2019} The first theory for a quantum spin liquid was the resonant valence bond (RVB) proposed by P. W. Anderson within the context of the spin-$\frac{1}{2}$ antiferromagnetic (AF) Heisenberg model on the triangular lattice\cite{Anderson73}. The RVB state is a linear superposition of different configurations of short range singlets where conventional spin-$1$  excitations can decay into pairs of spin-$\frac{1}{2}$ excitations due to the resonant (entangled) character of the RVB state. Even if subsequent works\cite{Huse1988, Bernu1992, Elstner1993, Bernu1994, Capriotti1999, White2007,Xie2020} demonstrate that the ground state of the triangular Heisenberg antiferromagnet has a 120$^{\circ}$ N\'eel structure with local magnetization $m=0.205$, it is believed that such reduction to 41$\%$ of the full moment is a signal of its proximity to a quantum melting point.\cite{Chubukov1994} In fact, numerical studies\cite{Kaneko14, Bishop15, McCulloch2016, Hu15, Zhu15, Iqbal16,  Oitmaa2020} show that a small amount of exchange interactions to next-nearest neighbors, the so-called $J_1-J_2$ model, induces a continuous transition to a quantum spin liquid state at just $J_2/J_1\approx 0.07$. In agreement with this idea, recent inelastic neutron scattering experiments in the effective $S=\frac{1}{2}$ triangular antiferromagnet 
Ba$_3$CoSb$_2$O$_9$\cite{Shirata2012, Susuki2013}, show an unusual extended and structured 
continuum\cite{Ma16, Ito2017, Kamiya, Coldea2020} that can not be accounted for by large-$S$ expansions, suggesting a significant amount of quantum fluctuations as expected in proximity to a quantum melting point \cite{Chernyshev09, Mourigal13, Ghioldi18}.

\begin{figure}[h]
\begin{center}
\includegraphics*[width=0.25\textwidth]{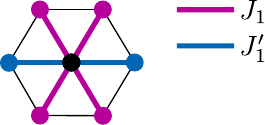}
\caption{Spatially anisotropic triangular lattice with $J_1$ and $J_1'$ nearest-neighbor interactions. $J_1=0$, $J_1'=0$, and $J_1=J'_1$, corresponds to chains, square and triangular AF, respectively.}
\label{fig1}
\end{center}
\end{figure}

Another related model that has been widely studied in the literature\cite{Trumper1999, Manuel1999, Yunoki2006, Heidarian09,Mischa2014, Ghorbani16} is the spatially anisotropic  $J_1-J_1'$ Heisenberg model on a triangular lattice, where $J_1$ runs along two directions and $J_1'$ along the other one (see Fig. \ref{fig1}). For $0 \le J_1/ J'_1 \le 1$ it corresponds to spin chains along $J'_1$ coupled through frustrating zigzag exchange $J_1$, interpolating between decoupled AF spin chains, $J_1/ J'_1=0$, and the spatially isotropic triangular antiferromagnet, $J_1/ J'_1=1$. In particular, this model has been proposed to describe the unusual excitation spectrum of the compound Cs$_2$CuCl$_4$ with $J_1/J'_1\approx 0.33$\cite{Coldea02, Coldea, Fjaerestad07, Starykh10} and the spiral features of the compound Cs$_2$CuBr$_4$ with $J_1/J'_1\approx 0.75$\cite{Zheng05, Fjaerestad07}. Recent variational Monte Carlo (VMC) calculations\cite{Ghorbani16} predict a quasi-one-dimensional gapless spin liquid for $J_1/J'_1\leq 0.6$ and incommensurate spiral phases for $J_1/J'_1\geq 0.6$ which are in agreement with the observed features of the compounds Cs$_2$CuCl$_4$ and Cs$_2$CuBr$_4$, respectively. Originally, the anomalous extended continuum observed in the excitation spectrum of Cs$_2$CuCl$_4$ was identified with the presence of 2D spinons,\cite{Coldea,Alicea2005,Yunoki2006} although it was recognized later that such a spinon continuum has a one dimensional (1D) character due to the dimensional reduction induced by the frustrating zigzag coupling\cite{Zheng, Kohno, Heidarian09, Starykh10}. This one-dimensionalization phenomenon has also been found in spin-$1$ systems.\cite{Gonzalez17, Hembacher18, Abdeldaim19} On the other hand, for $0 \le J'_1/ J_1 \le 1$, the model interpolates between the square AF, $J'_1/ J_1 =0$, and the triangular AF, $J'_1/ J_1 = 1$. VMC predicts a transition at $J'_1/J_1\approx 0.7$ from a collinear N\'eel phase to incommensurate spiral phases; while in the range $0.7 \le J'_1/ J_1 \le 0.8$ the $Z_2$ gapless spin liquid phase has very similar energies to the spiral ones\cite{Ghorbani16}. 

Motivated by the small amount of second neighbor exchange interaction ($J_2$) needed to stabilized 2D spin liquid 
phases\cite{Kaneko14, Bishop15, McCulloch2016, Hu15, Zhu15, Iqbal16,  Oitmaa2020} and the unusual effects of the spatial anisotropy\cite{Trumper1999, Manuel1999, Coldea02, Coldea, Yunoki2006, Fjaerestad07, Heidarian09, Starykh10, Mischa2014, Ghorbani16} ($J'_1$), in this paper we investigate the phase diagram of the spatially anisotropic $J_1-J_2$  model on triangular lattices (see Fig. \ref{fig2}). We use the Schwinger boson (SB) theory\cite{Arovas88, Auerbach1994} to compute the quantum phase diagram of the model up to Gaussian order\cite{Trumper97, Ghioldi18} ($1/\mathcal{N}$ correction)  where $\mathcal{N}$ is the flavor number of SB. At this level of calculation certain entanglement effects are taken into account through the fluctuations of the emergent gauge fields.\cite{Read1991} The mean-field phase diagram of this model was already computed by \textcite{Merino14}, finding, N\'eel, collinear, spiral, and spin liquid regions. However, it is well known that at the mean field level the magnetic ordering is overestimated and the magnetic excitations are not the physical ones. Therefore, the inclusion of Gaussian fluctuations is imperative.\cite{Trumper97, Ghioldi18} In fact, at Gaussian level the whole phase diagram is strongly renormalized with respect to the mean-field approach (see Fig. \ref{fig4} and Fig. \ref{fig5}). In particular, for all values of $J_2$ we find that quantum fluctuations reinforce the N\'eel phase; while the stability of the long range collinear and spiral regions are strongly reduced along with the appearance of $0$-flux and nematic spin liquids regions in between, corresponding to its short range counterparts. The good agreement between the Schwinger boson theory and the variational Monte Carlo predictions\cite{Ghorbani16} along the line $J_2=0$ gives a strong support to our results. Remarkably, around the isotropic line $J'_1=J_1$, the stability of the spirals is so weak, that our results seem to recover the transition to the spin liquid phase at the expected value $J_2/J_1\approx0.07$, found by the most sophisticated numerical methods.\cite{Hu15, Zhu15, Iqbal16, Oitmaa2020}        

In section II we present the spatially anisotropic $J_1-J_2$ Heisenberg model with next-nearest neighbors exchange interactions on the triangular lattice along with the well-known limits it covers and its corresponding classical phase diagram. In section III we present the main steps to compute the Gaussian corrections within the Schwinger boson theory. In section IV we present the Gaussian corrected phase diagram and compare it with the mean field one. In section V we close with the conclusions.     

\section{Spatially anisotropic $J_1$-$J_2$  Heisenberg model}

\begin{figure}[h]
\begin{center}
\includegraphics*[width=0.3\textwidth]{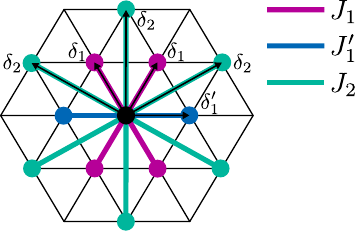}
\caption{Spatially anisotropic triangular lattice with next-nearest-neighbor interactions. $J_1$ runs along two triangular directions, $J_1'$ runs along the remaining one, and $J_2$ connects all next-nearest-neighbors on a triangular lattice (only the interactions to the center-site are shown for the sake of clarity).}
\label{fig2}
\end{center}
\end{figure}

We focus on the spatially anisotropic triangular spin-1/2 antiferromagnetic Heisenberg model with next-nearest-neighbor interactions, whose Hamiltonian can be written as
\begin{equation}
\mathcal{H} = J_{1} \sum_{\langle i,j\rangle} \mathbf{S}_{i} \cdot \mathbf{S}_{j} + J_{1}' \sum_{\langle i,j\rangle'} \mathbf{S}_{i} \cdot \mathbf{S}_{j} + J_{2} \sum_{[ k,l]} \mathbf{S}_k \cdot \mathbf{S}_{l},
\label{ham}
\end{equation}
\noindent where $ \mathbf{S}_{i}$ are the quantum spin-$\frac{1}{2}$ operators and all exchange interactions are positive. The sum ${\langle i,j\rangle}$ indicates nearest neighbors along two directions $\bm{\delta}_1$, ${\langle i,j\rangle'}$ indicates nearest neighbors along the remaining direction $\bm{\delta}'_1$, and $[ k,l]$ runs along next nearest neighbors $\bm{\delta}_2$ (see Fig. \ref{fig2}).

This Hamiltonian includes some well-known limits: for $J_2=J'_1 =0$ it reduces to the unfrustrated square-lattice antiferromagnetic Heisenberg model, which exhibits a N\'eel order in the thermodynamic limit; whereas for $J_2=0$ and $J'_1 = J_1$ it reduces to the triangular-lattice antiferromagnetic Heisenberg model which exhibits a commensurate 120$^\circ$ N\'eel order with three sublattices. Therefore, when varying $J'_1$ from 0 to $J_1$, our model Hamiltonian interpolates between the square and triangular lattices, respectively. For $J_1/J'_1 = 0$ the system becomes a set of decoupled one-dimensional gapless spin-$\frac{1}{2}$ chains with quasi long range magnetic order. Then, when increasing $J'_1$ from $J'_1=J_1$ we can interpolate between the 2D triangular lattice and a set of one-dimensional decoupled chains. Another important parameter space of the Hamiltonian is along the isotropic line, $J'_1=J_1$, with varying $J_2$, corresponding to the so-called $J_1-J_2$ Heisenberg model on the triangular lattice.\\

\begin{figure}[t]
\begin{center}
\includegraphics*[width=0.45\textwidth]{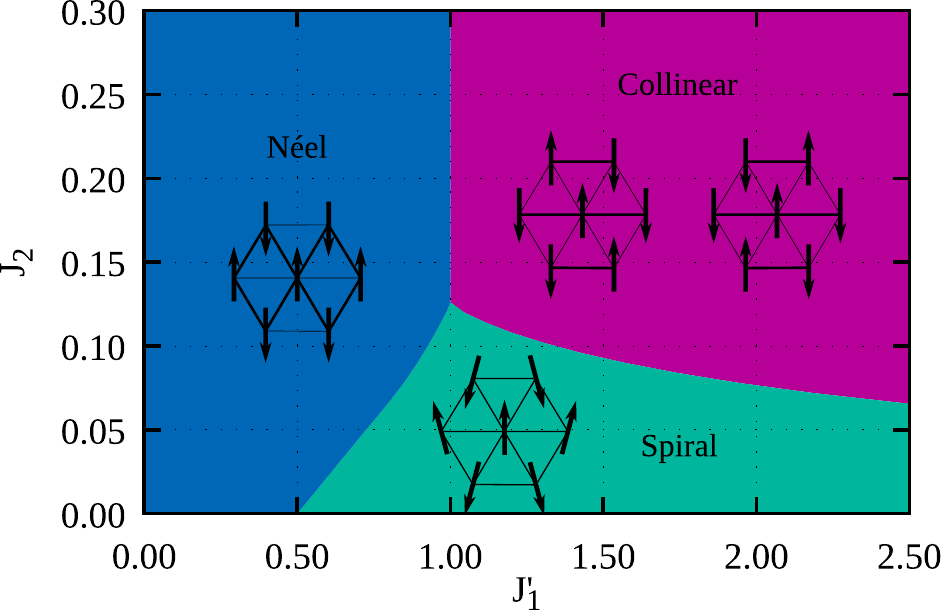}
\caption{Classical phase diagram of the Hamiltonian \ref{ham}. The N\'eel phase, $\mathbf{Q}=(0,\frac{2\pi}{\sqrt{3}})$  is colored in blue, the collinear phase, $(\pi,\frac{\pi}{\sqrt{3}})$, in purple, and the spiral phases, $(Q,\frac{2\pi}{\sqrt{3}})$ in green.}
\label{fig3}
\end{center}
\end{figure}

In Fig. \ref{fig3} is shown the classical phase diagram of Hamiltonian (\ref{ham}), obtained by minimizing the classical energy
\begin{equation}
E = S^2 \sum_{ij} J_{ij} \cos\left(\mathbf{Q}\cdot \bm{\delta}_{ij} \right),
\end{equation}
where the exchange $J_{ij}$ takes finite values $J_1, J'_1,$ and $J_2$ only along the directions $\bm{\delta}=\bm{\delta}_1,\bm{\delta}'_1$, $\bm{\delta}_2$ needed to build the system shown in Fig. \ref{fig2},
 and $J_1$ is taken as energy unit. The advantage of the present lattice is that all magnetic phases can be classified by a unique magnetic wave vector $\bf Q$. The phase diagram exhibits three different magnetic orders: N\'eel order signalled by $\mathbf{Q}=(0,\frac{2\pi}{\sqrt{3}})$ on the triangular Brillouin zone (blue), collinear magnetic order $(\pi,\frac{\pi}{\sqrt{3}})$ (purple), and incommensurate magnetic order $(Q,\frac{2\pi}{\sqrt{3}})$ (green). Along the spatially isotropic line $J'_1\!\! = \!\!J_1$ the N\'eel and collinear phases are the same, while the  spirals turn out commensurated of $120^\circ$ order $(\frac{2\pi}{3},\frac{2\pi}{\sqrt{3}})$. Furthermore, the transition between the spiral and N\'eel orders is continuous.

For spin $S=\frac{1}{2}$, the classical phase diagram is expected to change due to  quantum fluctuations, enhanced by frustrating interactions. For example, as discussed in the introduction, along the two lines characterized by $J'_1\! \!=\!\!J_1$, and $J_2\!\!=\!\!0$, numerical methods such as density-matrix renormalization group algorithms and variational Monte Carlo have predicted the existence of quantum spin liquid phases somewhere between the ordered phases.\cite{Hu15, Zhu15, Iqbal16, Oitmaa2020} The problem of mapping the whole quantum phase diagram has only been carried out recently within the Schwinger boson mean-field theory\cite{Merino14}, which we will discuss in the next sections.
Given the variety of frustrating interactions between different neighbors and the several ordered and disordered phases expected, exploring the whole phase diagram is not an easy task. The most standard exact diagonalization and density-matrix methods present an increasing computational cost when increasing system sizes and scaling towards the thermodynamic limit (further increased when including interactions between distant neighbors); while quantum Monte Carlo method suffers the sign problem. Consequently, to carry on this task we will use the Schwinger boson theory\cite{Auerbach1994} at the Gaussian correction level, or 1/$\mathcal{N}$ corrections, where $\mathcal{N}$ is the number of flavors of the Schwinger bosons. We have recently developed a detailed analysis of the theory beyond the mean-field approach.\cite{Ghioldi18} This theory has proven to give good quantitative results for static and dynamic properties, and its ability to describe both, ordered (commensurate and incommensurate) and disordered phases, allows us to compute a complete and reliable  phase diagram.\cite{Read1991, Ceccatto93, Trumper97, Manuel1999, Wang2006, Bauer17}

\section{Schwinger boson theory}

In the Schwinger boson theory\cite{Auerbach1994, Arovas88} the spin operator is represented in terms of bosonic spinor operators as  $
{\bf S}_i= \frac{1}{2}{\bf b}^{\dagger}_i \vec{\sigma} \;{\bf b}_i$,
where $\vec{\sigma}$ is the vector of Pauli matrices and ${\bf b}^{\dagger}_i =(\hat{b}^{\dagger}_{i\uparrow}; \hat{b}^{\dagger}_{i\downarrow})$ is the spinor
of SB's $\hat{b}_{\uparrow}$ and $\hat{b}_{\downarrow}$. To satisfy the spin algebra, a local constraint over the number of bosons per site $\sum_{\sigma}\hat{b}^{\dagger}_{i\sigma}\hat{b}_{i\sigma}=2S$ has to be imposed. 
In this representation, the Heisenberg interaction of equation (\ref{ham}) can be re-written in terms of SU(2) invariant link operators as  
\cite{Ceccatto93, Flint09, Ghioldi18}
\begin{equation}
H = \sum_{i,j} J_{ij} \left( :\hat{B}_{ij}^\dagger \hat{B}_{ij} : - \hat{A}_{ij}^\dagger \hat{A}_{ij} \right),
\label{hamsch}
\end{equation}
where $\hat{B}_{ij}^\dagger=\frac{1}{2} \sum_\sigma \hat{b}_{i\sigma}^\dagger \hat{b}_{j\sigma}$ and $\hat{A}_{ij}=\frac{1}{2} \sum_\sigma \sigma \hat{b}_{i\sigma} \hat{b}_{j\bar{\sigma}}$ are the building blocks of the theory. 
\\

Using coherent states of SB the partition function takes the form\cite{Arovas88,Ghioldi18} 
\begin{multline}
Z =  \int D\lambda [D\overline{b}Db]\ e^{-\int_0^\beta d\tau \left[ \sum_{i,\sigma} \overline{b}_{i\sigma}^\tau \partial_\tau b_{i\sigma}^\tau + H(\overline{b},b) \right] } \\
 \times e^{-\int_0^\beta d\tau \left[ i \sum_i \lambda_i^\tau (\sum_\sigma \overline{b}_{i\sigma}^\tau b_{i\sigma}^\tau -2S )\right] } ,
\end{multline}
where the $\lambda$ field is added to ensure the local constraint and the integrating measures are $[D\overline{b}D{b}]=\Pi \frac{d\overline{b}^{\tau}_{i\sigma}d{b}^{\tau}_{i\sigma}}{2\pi i}$ and  $D\lambda =\Pi \frac{d\lambda^{\tau}_i}{2\pi i}$. To decouple the $BB$ and $AA$ terms of the Hamiltonian, two types of Hubbard-Stratonovich fields, ${W}^A$ and $W^B$ are introduced. Then, the integrals in  $\overline{b}$ and $b$ can be carried out, leading to\cite{Ghioldi18}
\begin{equation}
Z = \int D\overline{W}DW D\lambda \ e^{-S_{\text{eff}}(\overline{W},W,\lambda)},
\label{effective}
\end{equation}
where $\overline{W}$ and $W$ denote the complex fields $\overline{W}^A$,  $\overline{W}^B$, $W^A$, and ${W}^B$, respectively, with the effective action $S_{\text{eff}}$ given by
\begin{multline}
S_{\text{eff}}\! =\!\! \int^{\beta}_0 \!\!\! \!d\tau (\sum_{i,j,\mu}\!\! J_{ij} \overline{W}_{ij}^{\mu,\tau} W_{ij}^{\mu,\tau} \!\!- i2S \sum_i \lambda_i^{\tau}) - \ln Z_{\text{bos}},
\end{multline}
where $\mu$ sums over the fields $A$ and $B$; and $Z_{\text{bos}}$ is the bosonic partition function 
\begin{equation}
 Z_{\text{bos}}= \int [D\overline{b}D{b}] e^{-S_{\text{bos}}(\overline{b},{b})}
\end{equation}
\noindent that integrates over the quadratic bosonic action given by\cite{Ghioldi18}
\begin{equation}
S_{\text{bos}} = \int^{\beta}_{0} \!\!\! d\tau \sum_{i,j} \vec{b}_i^{\tau\dagger} \mathcal{M}^{\tau}_{i,j} \vec{b}^{\tau}_{j}.
\end{equation}

Next, the effective action $ S_{\text{eff}}(\overline{W},W,\lambda)$ in Eq. (\ref{effective}) is expanded up to second order around the saddle-point solution of the fields,\cite{Gonzalez17, Ghioldi18}
\begin{equation}
S_{\text{eff}}(\overline{W},W,\lambda) \simeq  S^{(0)}_\text{eff} + \frac{1}{2} \sum_{\alpha_1,\alpha_2} \Delta \vec{\phi}_{\alpha_1}^\dagger S^{(2)}_{\alpha_1,\alpha_2}  \Delta \vec{\phi}_{\alpha_2}, 
\end{equation}
\noindent where $S^{(0)}_\text{eff}=S_{\text{eff}}(\overline{W}_{\text{sp}},W_{\text{sp}},\lambda_{\text{sp}})$ is the saddle-point effective action, $S^{(2)}= \frac{\partial^2 S_\text{eff}}{\partial \vec{\phi}^\dagger \partial \vec{\phi}}\big|_{\text{sp}}$ is the fluctuation matrix evaluated  at the saddle-point solution, $\alpha$ denotes momentum, frequency, and neighbor index; and   $\Delta \vec{\phi}^\dagger$ are the fluctuations of the fields around the saddle-point solution, defined by $\Delta \vec{\phi}^\dagger\!\! =\! \vec{\phi}^\dagger\! - \vec{\phi}_\text{sp}^\dagger $ 
where $\vec{\phi}\! =\! (W^B,\overline{W}^B,W^A,\overline{W}^A,
\lambda)^{\dagger}$, and $\vec{\phi}_\text{sp}^\dagger$ is the saddle-point solution that fulfills the condition $S^{(1)}=\frac{\partial S_{\text{eff}}}{\partial \vec{\phi}} =0$. Choosing a static and homogeneous \textit{ansatz}, the saddle-point solution can be related to the real mean-field parameters  ${A}_{\bm{\delta}}$, ${B}_{\bm{\delta}}$, and $\lambda$ as follows 
\begin{equation}
\begin{array}{c}
{W}_{\bm{\delta}}^B \Big|_\text{sp} = -\sqrt{N\beta} B_{\bm{\delta}}, \ \ \ \ \ \ \ \ \ \  {W}_{\bm{\delta}}^A \Big|_\text{sp} = i\sqrt{N\beta} A_{\bm{\delta}},\\
 \lambda \Big|_\text{sp}=i \lambda,
\end{array}
\end{equation}
\noindent where $B_{\bm{\delta}}= \langle \hat{B}_{\bm{\delta}} \rangle$ and $A_{\bm{\delta}}= \langle \hat{A}_{\bm{\delta}} \rangle$. The saddle-point condition leads to the self-consistent equations, which have the usual zero-temperature form of the SB mean-field theory\cite{Ceccatto93,Mezio11}
\begin{equation}
{A}_{\bm{\delta}} = \frac{1}{2N} \sum_{\bf k} \frac{\gamma_{\bf k}^A}{\varepsilon_{\bf k}} \sin({\bf k}\cdot {\bm{\delta}}) 
\label{A},
\end{equation}
\begin{equation}
{B}_{\bm{\delta}} = \frac{1}{2N} \sum_{\bf k} \frac{\gamma_{\bf k}^B + \lambda}{\varepsilon_{\bf k}} \cos({\bf k}\cdot {\bm{\delta}}),
\label{B}
\end{equation}
\begin{equation}
S+\frac{1}{2} = \frac{1}{2N} \sum_{\bf k} \frac{\gamma_{\bf k}^B + \lambda}{\varepsilon_{\bf k}}.
\label{const}
\end{equation}
where the free spin-$\frac{1}{2}$ spinon dispersion relation is obtained by diagonalizing the mean-field Hamiltonian:\cite{Mezio11}
\begin{equation} 
\varepsilon_{\bf k} = \sqrt{\left(\gamma_{\bf k}^B+\lambda \right)^2-\left(\gamma_{\bf k}^A\right)^2},
\end{equation}
with $\gamma_{\bf k}^B \!\!=\!\!\ \sum_{\bm{\delta}} J_{\bm{\delta}} B_{\bm{\delta}} \cos({\bf k}\cdot {\bm{\delta}})$,  $\gamma_{\bf k}^A\! =\! \sum_{\bm{\delta}} J_{\bm{\delta}} {A}_{\bm{\delta}} \sin({\bf k}\cdot {\bm{\delta}})$, and the sums go over ${\bm {\delta}}\!\!=\!\!{\bm{\delta}_1}, {\bm{\delta}'_1}, {\bm{\delta}_2}$. Notice that the physical spin-$1$ excitations at the mean field level involve a continuum of two free spinon excitations. We have recently shown that when the ground state is magnetically ordered the Gaussian corrections induce the expected collective magnon excitations as two-spinon bound states.\cite {Ghioldi18, Zhang}\\

For any finite lattice, the mean field ground state has a  singlet nature due to the rotational invariant character of the operators $\hat{A}_{ij}$ and $\hat{B}_{ij}$. \cite{Mezio11}  Nevertheless, as the system size $N$ increases the spinon gap may behave as $\varepsilon_{\pm{{\bf Q}_0}/{2}}\sim 1/N$ for a given ${\bf Q}_0$. In the thermodynamic limit, these zero modes can be treated as Bose condensates that lead to the putative rupture of the SU(2) symmetry.\cite{Hirsch89, Sarker, Chandra90} In this case the local magnetization $m_{\text{sp}}({\bf Q}_0)$ can be extracted from the singular part of Eqs. (\ref{A})-(\ref{const}) while ${\bf Q}_0$ is the magnetic wave vector of the long range order structure. The Gaussian corrections to $m_{\text{sp}}({\bf Q}_0)$, however, require a much more involved calculation. Namely, introducing an infinitesimal magnetic field along the  local magnetic order that is sent to zero after the thermodynamic limit is carried on.\cite{Ghioldi18} Alternatively, in order to evaluate the presence of long range order at Gaussian level we compute the magnetic spin stiffness. The advantage is that the spin stiffness can be computed on finite systems, allowing us an appropriate size scaling study.\cite{Trumper97, Manueltriang, Manuel1999, Gonzalez17}
The procedure consists of solving the equations (\ref{A})-(\ref{const}) with twisted boundary conditions in such a way that the saddle-point solution corresponds to a magnetic structure slightly twisted by $\Delta {\bf Q}$ from ${\bf Q}_0$. So now the ground-state energy is a function of the twisted wave vector ${\bf Q} = {\bf Q}_0+ \Delta {\bf Q} $ through the mean-field parameters as    
\begin{equation}
E_\text{SP}({\bf Q})= N \sum_{\bm{\delta}} B_{\bm{\delta}}^2({\bf Q}) - A_{\bm{\delta}}^2({\bf Q}),
\end{equation}
whose spin stiffness is obtained by 
\begin{equation}
\rho_\text{SP}= \frac{\partial^2 E_\text{SP}({\bf Q})}{\partial {\bf Q}^2}\Big|_{{\bf Q}_\text{0}},
\label{stiff}
\end{equation}
where the second order derivative is evaluated at the local minimum ${{\bf Q}_\text{0}}$ of the ground-state energy $E_\text{SP}({\bf Q})$.\\

 The other advantage of the spin stiffness is that the Gaussian corrections can be easily calculated by replacing the Gaussian corrected ground-state energy in Eq. (\ref{stiff}). 
This requires integrating the Gaussian fluctuations of the Hubbard-Stratonovich fields which are the gauge fields of the effective partition function  
\begin{equation}
Z \simeq e^{-S^{(0)}_\text{eff}} \times  \int D\vec{\phi}^{\dagger} D\vec{\phi}  \ e^{- \frac{1}{2} \Delta \vec{\phi}^\dagger S^{(2)}  \Delta \vec{\phi}}.
\end{equation}

\noindent However, due to the rupture of the local gauge symmetry of the saddle-point solution, the fluctuation matrix $S^{(2)}$ has infinite zero modes related to the gauge fluctuations that lead to divergences. To avoid them it is used the Fadeev-Popov trick which restricts the integration to field fluctuations orthogonal to the gauge orbit \cite{Trumper97, Ghioldi18}. Alternatively, one can obtain exactly the same result by truncating the $\lambda$ field column and row of  $S^{(2)}$ (resulting in truncated fluctuation matrix $S_\text{tr}^{(2)}$)\cite{Ghioldi18}. In this latter case, the Gaussian correction to the ground-state energy (zero temperature) gives   
\begin{equation}
 E^{(2)}=-\frac{1}{4\pi N} \int^{\infty}_{-\infty}d\omega \sum_{\bf k} \ln \left[ \frac{1}{\det S_{\text{tr}}^{(2)}({\bf k},\omega)} \right].
\end{equation}
Then, we can calculate the ground-state energy at Gaussian order for any twisted boundary condition as $E_\text{FL}({\bf Q})=E_\text{SP}({\bf Q}) + E^{(2)}({\bf Q})$, and therefore also the spin stiffness 
\begin{equation}
 \rho_\text{FL}=\frac{\partial^2 E_{\text {FL}}({\bf Q})}{\partial {\bf Q}^2}\Big|_{{\bf Q}_\text{min}^\text{FL}},
\end{equation}
where ${\bf Q}_\text{min}^\text{FL}$ is the local minimum of $E_\text{FL}({\bf Q})$, that can be different from ${\bf Q}_\text{0}$ of the saddle-point solution.\cite{Manuel1999}
\\

\section{Phase diagram}

In general we use clusters that respect the symmetry of the triangular lattice in the thermodynamic limit, of the form  $N=  3\times L \times L $, with periodic boundary conditions.\cite{Bernu1994} For the mean field solutions we practically have no restriction for the size scaling of the spin stiffness; while for the Gaussian corrections we use system sizes up to $N=1200$ sites.

\begin{figure}[t]
\begin{center}
\includegraphics*[width=0.45\textwidth]{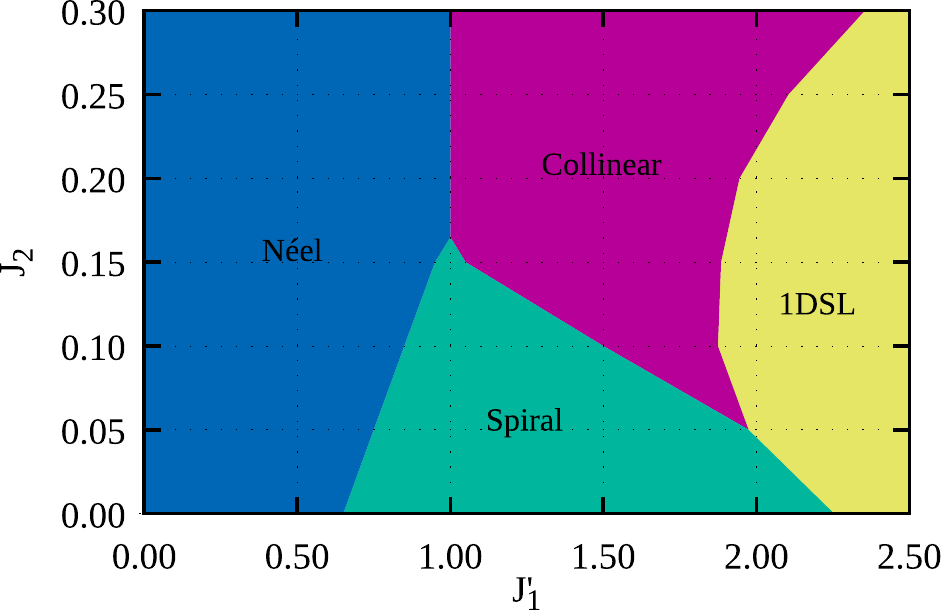}
\caption{Schwinger boson mean field phase diagram for the spatially anisotropic  $J_1-J_2$ model of equation (\ref{ham}). The labels are as in Fig. \ref{fig3}, except 1DSL, which corresponds to the quantum spin liquid with 1D nature (see text). }
\label{fig4}
\end{center}
\end{figure}

\begin{figure}[t]
\begin{center}
\includegraphics*[width=0.45\textwidth]{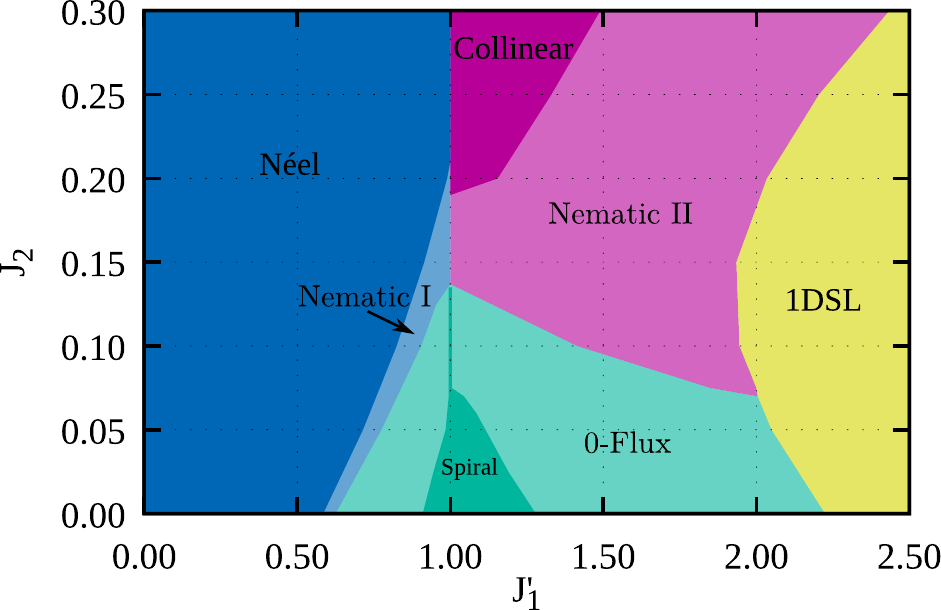}
\caption{Gaussian corrected mean field phase diagram for the spatially anisotropic  $J_1-J_2$ model of equation (\ref{ham}). The labels are as in Fig. \ref{fig4}. The $0$-flux, nematic I, and nematic II corresponds to the short range order counterparts of the spiral, N\'eel, and collinear phases, respectively. See table I for its representative parameter structure.}
\label{fig5}
\end{center}
\end{figure}

To get the mean field phase diagram we solve the self-consistent equations (\ref{A})-(\ref{const}) by plugging in the different classical ansatzs $A_{\bm{\delta}}=S \sin({\bf Q} \cdot {\bm{\delta}}/2)$ and $B_{\bm{\delta}}=S \cos({\bf Q} \cdot {\bm{\delta}}/2)$,  where ${\bf Q}=(0, \frac{2\pi}{\sqrt{3}})$, ${\bf Q}=(\pi, \frac{\pi}{\sqrt{3}})$, and ${\bf Q}=(Q,\frac{2\pi}{\sqrt{3}})$ correspond to N\'eel, collinear, and spiral phases, respectively. The resulting mean field phase diagram is shown in Fig. \ref{fig4} where there are two main points to stress. On one hand, the quantum fluctuations enhance the stability of the spiral phases with respect to N\'eel and collinear ones. This effect is mostly observed  along the spatially isotropic line, $J_1\!\!=\!\!J'_1$. Here, the N\'eel and collinear phases coincide again; while the transition between the spiral and the N\'eel phase is continuous as in the classical case. 
On the other hand, for $J'_1 \gtrsim 2$, the collinear and spiral phases melt with the appearance of a large quantum disordered region (yellow region on the right of the diagram). Along this boundary the spin stiffness vanishes. Notice that the same mean field phase diagram was obtained by \textcite{Merino14}  using the local magnetization instead of the spin stiffness, as order parameter. In addition we have found that the quantum spin liquid region has a one dimensional character, a feature that was overlooked. Namely, besides the vanishing of the spin stiffness (local magnetization) the mean field solution corresponds to a collection of decoupled spin chains with finite $A_{{\bm{\delta}}'_1}$ along ${\bm{\delta}}'_1$, whereas $A_{{\bm{\delta}}_1}\!=\!B_{{\bm{\delta}}_1}\!=\!0$ along ${\bm{\delta}}_1$. For this reason, in Fig. \ref{fig5}, the spin liquid region has been called 1D spin liquid (1DSL).   

\begin{table}
\begin{tabular}[c]{ccccccccc}
\hline
\hline
                     &&  $\;\;\;\;\;\;\;\;\;$ $0$-flux   $\;\;\;\;\;\;\;\;\;\;\;\;$    && nematic I  $\;\;\;\;\;\;\;\;\;$&& nematic II  \\
\hline
\hline
$A_{{\bm{\delta}}_1}$            && -$(A,A',A')$      && -$(0,A,A)$       && $-(A,A',0)$\\
 $B_{{\bm{\delta}}_1}$            && $(B,$-$B',B')$       && $(B,0,0)$      && $(0,0,B)$    \\
  $A_{{\bm{\delta}}_2}$       && $(\pm A''\!,0,\pm A'')$     && -$(A',0,A')$    && $(0,$-$A''\!,A''')$  \\
  $B_{{\bm{\delta}}_2}$             &&$($-$B''\!,$-$B'''\!,B'')$      && $(0,$-$B',0)$  && $($-$B',0,0)$ \\
\tableline
\end{tabular}
\caption{Structure of the mean field parameters for the $0$-flux, nematic I, and nematic II spin liquid phases. The $\pm$ correspond to $J_1\!\!<\!\!J'_1$ and $J_1\!\!>\!\!J'_1$, respectively. Along the isotropic line, $J_1\!\!=\!\!J'_1$, nematic I and nematic II have equivalent structures.}
\label{table1}
\end{table}

The Gaussian corrected phase diagram is shown in Fig. \ref{fig5}. It is observed a strong reduction of all long range ordered regions, except the N\'eel phase, with respect to the mean field ones, accompanied by the emergence of its corresponding short range order (SRO) counterparts. Indeed, along the magnetically ordered boundaries the spin stiffness vanishes. The criteria to establish the stability of the SRO regions is the following: once long range order is lost, that is, the Gaussian corrected ground state energy $E_{FL} ({\bf Q})$ has not upward concavity as a function of ${\bf Q}$, we study the stability of each short range order regime by checking the positivity of the fluctuations matrix at ${\bf Q}_0$ (without twisted boundary conditions) through $\det S_{\text{tr}}^{(2)}({\bf k},\omega)$ (section III). Then, according to he projective symmetry group classification,\cite{Wen2002,Wang2006,Messio,Lu16} the SRO spiral solution corresponds to the $0$-flux spin liquid phase; whereas, both, SRO N\'eel and collinear solutions correspond to nematic I and nematic II spin liquid phases, respectively.\cite{Lu16, Bauer17} The structure of the mean field parameters for the $0$-flux, nematic I and nematic II spin liquid phases of the phase diagram (Fig. \ref{fig5}) is shown in table I.

\begin{figure}[t]
\begin{center}
\includegraphics*[width=0.45\textwidth]{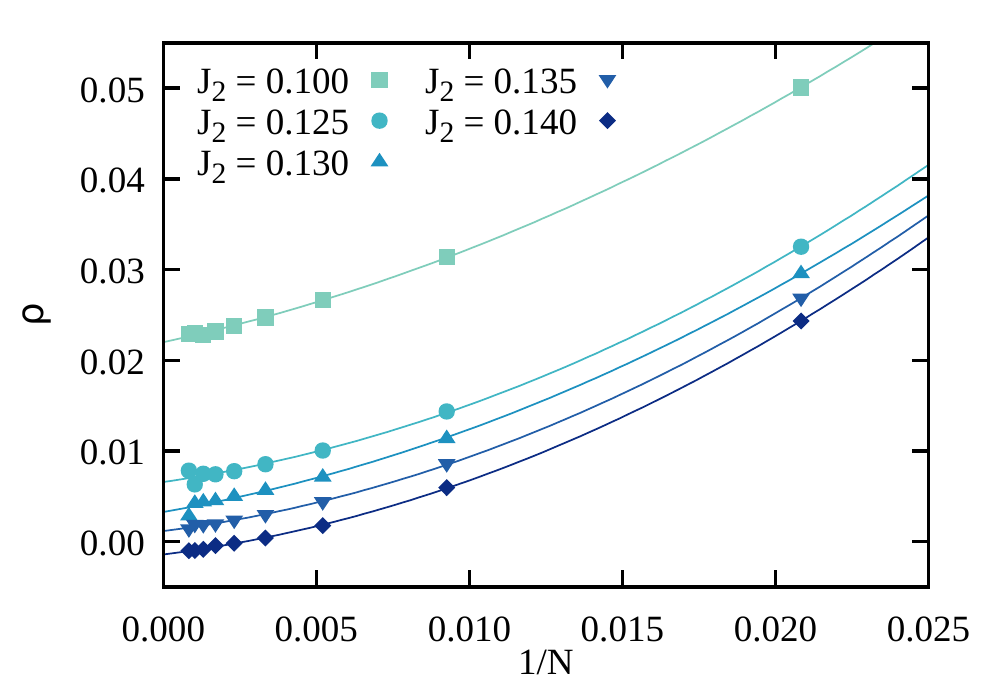}
\caption{Size scaling of the Gaussian corrected spin stiffness along the isotropic line $J_1=J'_1$  ($J_1-J_2$ model).}
\label{fig6}
\end{center}
\end{figure}

 On the other hand, the boundaries between such spin liquid phases have been located by using energy arguments. Remarkably, in the neighborhood of the isotropic line $J_1\!\!=\!\!J'_1$, in the range $0.07\lesssim J_2 \lesssim 0.14$, the long range incommensurate spirals are so fragile that only the commensurate $120^{\circ}$ N\'eel state survives. This can be seen in Fig. \ref{fig6} where the size scaling of the spin stiffness along the isotropic line, shows that the $120^{\circ}$ N\'eel state survives for $J_2\lesssim 0.14 $. The unexpected  instability of the incommensurate spirals is quite suggestive since if such fragility would extend to the commensurate spiral case the corresponding critical value would be $J_2/J_1 \approx 0.07$, which coincides with the quantum phase transition predicted by the more sophisticated numerical methods\cite{Hu15, Zhu15, Iqbal16,  Oitmaa2020} for the $J_1-J_2$ model on the triangular lattice. Furthermore, the competence among several SRO phases near $J_2/J_1\approx 0.14$  demonstrates the difficulties to discern the actual nature of the spin liquid phase in the $J_1-J_2$ model.

\begin{figure}[t]
\begin{center}
\includegraphics*[width=0.45\textwidth]{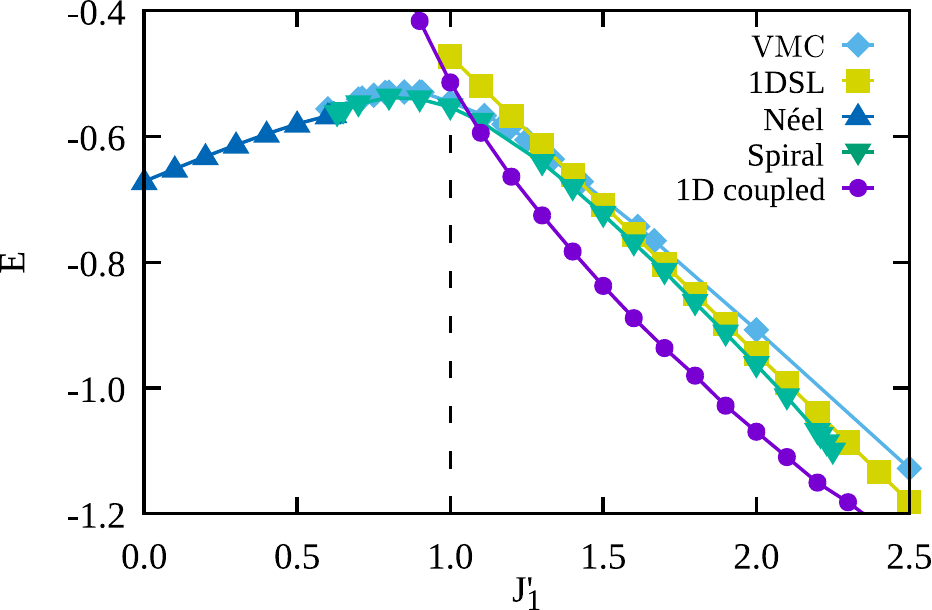}
\caption{ Comparison of the ground state energy per site between variational Monte Carlo and the Gaussian corrected Schwinger boson theory along the line $J_2=0$. Light blue diamonds are the variational Monte Carlo results taken from Ref. [\onlinecite{Ghorbani16}]. Green squares are for decouples spin chains, while violet circles correspond to weakly coupled spin chains (see text). }
\label{fig7}
\end{center}
\end{figure}
In order to study the validity of our results we concentrate along the line $J_2=0$ of the phase diagram where variational Monte Carlo results are available.\cite{Heidarian09, Ghorbani16} In Fig. \ref{fig7} it is compared the ground state energy per site predicted by variational Monte Carlo with the SB theory at the Gaussian level. Actually, the VMC results correspond to lattice sizes of $N=18 \times 18=324$ (square geometry) while the SB ones are for $N= 3\times12 \times 12  =432$ size (triangular geometry). The SB theory reproduces quite well the energies of N\'eel and spiral phases, but in the regime of weakly coupled chains the Gaussian corrected energies (violet circles) get worse with respect to VMC. We suspect that this behavior is related to the failure of the SB theory to recover the gapless nature of the spin-$\frac{1}{2}$ chains. Instead, for decoupled chains the Gaussian corrected energy (green square) shows a linear dependence with $J'_1$ which agrees better with the numerical results (turquoise diamond). In other words, once the spin-$\frac{1}{2}$ chains are coupled, at this approximation level, the SB theory does not capture properly the one-dimensionalization phenomenon.\cite{Zheng,Yunoki2006, Heidarian09,  Starykh10} For this reason, in order to locate  the boundary of the 1DSL region in the phase diagram (Fig. \ref{fig5}), we have used the results  corresponding to completely decoupled spin chains. Furthermore, the very similar energy values of the 1DSL and $0$-flux phase for an important range of $J'_1$ does not allow to discern precisely the boundary between them. So that, the actual boundary of the 1DSL region will be  surely modified.  \\

 \begin{figure}[h]
\begin{center}
\includegraphics*[width=0.45\textwidth]{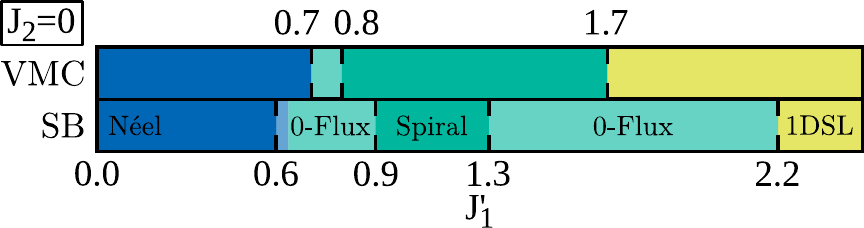}
\caption{Comparison between variational Monte Carlo taken from Ref. [\onlinecite{Ghorbani16}] and Gaussian corrected Schwinger boson theory, along the line $J_2=0$.}
\label{fig8}
\end{center}
\end{figure}

 In Fig. \ref{fig8} we compare our results with the prediction of the variational Monte Carlo\cite{Ghorbani16} along the line $J_2=0$. VMC predicts a $Z_2$ spin liquid with gapless nature between N\'eel and spiral phases. This state, however, has a very similar energy to the spiral one within the range $0.7 \le J'_1 \le 0.8$. On the other hand, at Gaussian level, the SB theory predicts the $0$-flux phase within the range 0.6$\lesssim J'_1\lesssim 0.9$. Given that the parameters $A_{\bm{\delta}}$ and $B_{\bm{\delta}}$ of the $0$-flux phase are non zero, it corresponds to a $Z_2$ spin liquid, but of gapped nature.\cite{Wang2006} Then, even if the spin liquid regions predicted by VMC $(0.7-0.8$) and SB $(0.6-0.9)$, are shown in the same colour in Fig. \ref{fig8}, they are not the same phase. Regarding the critical value between the 1D spin liquid and the $0$-flux, we believe that the 1D spin liquid phase will be more extended, as discussed above, shrinking the $0$-flux region, although it will probably not disappear as in VMC case. At this point, it is worth to stress that, besides of the fermionic representation for the spin operators, the VMC relies on the selected variational wave function; while the SB theory is based on the bosonic representation and relies on the link operators $\hat{A}_{ij}$ and $\hat{B}_{ij}$ by mean of which the Heisenberg interaction is expressed. The validity of both methods  in the whole parameter space should be more carefully investigated.\cite{Sorella2012}

\section{Conclusions}

We have computed the phase diagram of the spatially anisotropic spin-$\frac{1}{2}$ AF Heisenberg model on triangular lattice with next-nearest neighbor interactions. We have used the Schwinger boson theory up to Gaussian order. The phase diagram (Fig. \ref{fig5}) consists of an important region with long range N\'eel order and well reduced regions with long range collinear and spiral phases with respect to the mean field ones (Fig. \ref{fig4}). This reduction of the long range order regions is accompanied by the emergence of its short range counterparts, leaving an ample room for $0$-flux and nematic spin liquid phases. Unlike the SB mean field approach, which favors magnetically ordered phases, the Gaussian corrections stabilize spin liquid phases. Our results compare quite well with variational Monte Carlo along the line $J_2=0$, although the location of the boundary of the 1D spin liquid region (yellow region of the phase diagram) is not very reliable. This is probably related to the failure of the Schwinger boson theory to recover, at Gaussian level, the quasi-one dimensional regime. Remarkably, within the neighborhood of the isotropic line, $J_1=J'_1$, the incommensurate spirals are so fragile that only survives the commensurate $120^{\circ}$ N\'eel ones. This result is quite suggestive since if such a fragility were also for the commensurate ones the melting point would be at $J_2/J_1\approx 0.07$, which agrees with predictions of the more sophisticated numerical methods for the $J_1-J_2$ model.\cite{Kaneko14, Bishop15, McCulloch2016, Hu15, Zhu15, Iqbal16,  Oitmaa2020} Our study demonstrate the need to incorporate  Gaussian fluctuations above the SB mean field approach to obtain a very rich phase diagram. We hope that this work along with the study of other related models\cite{White2018} serve to guide the increasing search of triangular AF compounds with spin liquid behaviour.\cite{li2020}  \\

We remind that the Gaussian corrections to the mean field solution incorporate the fluctuation of the Hubbard-Stratonovich $W^A, W^B$ and $\lambda$ fields which are the gauge fields of the effective theory. Besides of changing the mean field ground state   it is important to point out that the fluctuations of the $\lambda$ field  improve the local constraint of the SB's, which is relaxed at the mean field level; while the $W^A$ and $W^B$ fluctuations mediate the interaction between spinons that at the mean field level are free. Recent computation of the Gaussian corrected  dynamical structure factor of the triangular Heisenberg model shows that in the magnetic excitation spectrum coexist an extended two-spinon continuum along with collective magnon excitations as two-spinon bound states.\cite{Ghioldi18, Zhang} Though it is out of the scope of the present work, we can conjecture that the magnetic spectrum of the long range order regions discussed above will have such coexistence of excitations with a relative spectral weight dependent of the frustration degree; while in the spin liquid phases the spectrum will be described by an extended continuum of spinon excitations.\cite{Becca} We leave this investigation for a future work.\\

After completing this work we came across with an exact diagonalization study performed in a related spatially anisotropic $J_1-J_2$ model on the triangular lattice whose phase diagram shows similar characteristics to our results.\cite{Wu2020}       \\

This work was supported by CONICET under Grants 423 No. 364 (PIP2015). 

\bibliography{Papers}

\begin{thebibliography}{71}%
\makeatletter
\providecommand \@ifxundefined [1]{%
 \@ifx{#1\undefined}
}%
\providecommand \@ifnum [1]{%
 \ifnum #1\expandafter \@firstoftwo
 \else \expandafter \@secondoftwo
 \fi
}%
\providecommand \@ifx [1]{%
 \ifx #1\expandafter \@firstoftwo
 \else \expandafter \@secondoftwo
 \fi
}%
\providecommand \natexlab [1]{#1}%
\providecommand \enquote  [1]{``#1''}%
\providecommand \bibnamefont  [1]{#1}%
\providecommand \bibfnamefont [1]{#1}%
\providecommand \citenamefont [1]{#1}%
\providecommand \href@noop [0]{\@secondoftwo}%
\providecommand \href [0]{\begingroup \@sanitize@url \@href}%
\providecommand \@href[1]{\@@startlink{#1}\@@href}%
\providecommand \@@href[1]{\endgroup#1\@@endlink}%
\providecommand \@sanitize@url [0]{\catcode `\\12\catcode `\$12\catcode
  `\&12\catcode `\#12\catcode `\^12\catcode `\_12\catcode `\%12\relax}%
\providecommand \@@startlink[1]{}%
\providecommand \@@endlink[0]{}%
\providecommand \url  [0]{\begingroup\@sanitize@url \@url }%
\providecommand \@url [1]{\endgroup\@href {#1}{\urlprefix }}%
\providecommand \urlprefix  [0]{URL }%
\providecommand \Eprint [0]{\href }%
\providecommand \doibase [0]{http://dx.doi.org/}%
\providecommand \selectlanguage [0]{\@gobble}%
\providecommand \bibinfo  [0]{\@secondoftwo}%
\providecommand \bibfield  [0]{\@secondoftwo}%
\providecommand \translation [1]{[#1]}%
\providecommand \BibitemOpen [0]{}%
\providecommand \bibitemStop [0]{}%
\providecommand \bibitemNoStop [0]{.\EOS\space}%
\providecommand \EOS [0]{\spacefactor3000\relax}%
\providecommand \BibitemShut  [1]{\csname bibitem#1\endcsname}%
\let\auto@bib@innerbib\@empty
\bibitem [{\citenamefont {Sachdev}(2008)}]{Sachdev2008}%
  \BibitemOpen
  \bibfield  {author} {\bibinfo {author} {\bibfnamefont {S.}~\bibnamefont
  {Sachdev}},\ }\href {\doibase 10.1038/nphys894} {\bibfield  {journal}
  {\bibinfo  {journal} {Nat. Phys.}\ }\textbf {\bibinfo {volume} {4}},\
  \bibinfo {pages} {173} (\bibinfo {year} {2008})}\BibitemShut {NoStop}%
\bibitem [{\citenamefont {Normand}(2009)}]{Normand2009}%
  \BibitemOpen
  \bibfield  {author} {\bibinfo {author} {\bibfnamefont {B.}~\bibnamefont
  {Normand}},\ }\href {\doibase 10.1080/00107510902850361} {\bibfield
  {journal} {\bibinfo  {journal} {Contemp. Phys.}\ }\textbf {\bibinfo {volume}
  {50}},\ \bibinfo {pages} {533} (\bibinfo {year} {2009})}\BibitemShut
  {NoStop}%
\bibitem [{\citenamefont {Savary}\ and\ \citenamefont
  {Balents}(2016)}]{Savary16}%
  \BibitemOpen
  \bibfield  {author} {\bibinfo {author} {\bibfnamefont {L.}~\bibnamefont
  {Savary}}\ and\ \bibinfo {author} {\bibfnamefont {L.}~\bibnamefont
  {Balents}},\ }\href {\doibase 10.1088/0034-4885/80/1/016502} {\bibfield
  {journal} {\bibinfo  {journal} {Rep. Prog. Phys.}\ }\textbf {\bibinfo
  {volume} {80}},\ \bibinfo {pages} {016502} (\bibinfo {year}
  {2016})}\BibitemShut {NoStop}%
\bibitem [{\citenamefont {Zhou}\ \emph {et~al.}(2017)\citenamefont {Zhou},
  \citenamefont {Kanoda},\ and\ \citenamefont {Ng}}]{Zhou2017}%
  \BibitemOpen
  \bibfield  {author} {\bibinfo {author} {\bibfnamefont {Y.}~\bibnamefont
  {Zhou}}, \bibinfo {author} {\bibfnamefont {K.}~\bibnamefont {Kanoda}}, \ and\
  \bibinfo {author} {\bibfnamefont {T.-K.}\ \bibnamefont {Ng}},\ }\href
  {\doibase 10.1103/RevModPhys.89.025003} {\bibfield  {journal} {\bibinfo
  {journal} {Rev. Mod. Phys.}\ }\textbf {\bibinfo {volume} {89}},\ \bibinfo
  {pages} {025003} (\bibinfo {year} {2017})}\BibitemShut {NoStop}%
\bibitem [{\citenamefont {Broholm}\ \emph {et~al.}(2020)\citenamefont
  {Broholm}, \citenamefont {Cava}, \citenamefont {Kivelson}, \citenamefont
  {Nocera}, \citenamefont {Norman},\ and\ \citenamefont
  {Senthil}}]{Broholm2020}%
  \BibitemOpen
  \bibfield  {author} {\bibinfo {author} {\bibfnamefont {C.}~\bibnamefont
  {Broholm}}, \bibinfo {author} {\bibfnamefont {R.~J.}\ \bibnamefont {Cava}},
  \bibinfo {author} {\bibfnamefont {S.~A.}\ \bibnamefont {Kivelson}}, \bibinfo
  {author} {\bibfnamefont {D.~G.}\ \bibnamefont {Nocera}}, \bibinfo {author}
  {\bibfnamefont {M.~R.}\ \bibnamefont {Norman}}, \ and\ \bibinfo {author}
  {\bibfnamefont {T.}~\bibnamefont {Senthil}},\ }\href {\doibase
  10.1126/science.aay0668} {\bibfield  {journal} {\bibinfo  {journal}
  {Science}\ }\textbf {\bibinfo {volume} {367}},\ \bibinfo {pages} {263}
  (\bibinfo {year} {2020})}\BibitemShut {NoStop}%
\bibitem [{\citenamefont {Wen}(2019)}]{Wen2019}%
  \BibitemOpen
  \bibfield  {author} {\bibinfo {author} {\bibfnamefont {X.-G.}\ \bibnamefont
  {Wen}},\ }\href {\doibase 10.1126/science.aal3099} {\bibfield  {journal}
  {\bibinfo  {journal} {Science}\ }\textbf {\bibinfo {volume} {363}},\ \bibinfo
  {pages} {834} (\bibinfo {year} {2019})}\BibitemShut {NoStop}%
\bibitem [{\citenamefont {Anderson}(1973)}]{Anderson73}%
  \BibitemOpen
  \bibfield  {author} {\bibinfo {author} {\bibfnamefont {P.~W.}\ \bibnamefont
  {Anderson}},\ }\href {\doibase https://doi.org/10.1016/0025-5408(73)90167-0}
  {\bibfield  {journal} {\bibinfo  {journal} {Mat. Res. Bull.}\ }\textbf
  {\bibinfo {volume} {8}},\ \bibinfo {pages} {153 } (\bibinfo {year}
  {1973})}\BibitemShut {NoStop}%
\bibitem [{\citenamefont {Huse}\ and\ \citenamefont {Elser}(1988)}]{Huse1988}%
  \BibitemOpen
  \bibfield  {author} {\bibinfo {author} {\bibfnamefont {D.~A.}\ \bibnamefont
  {Huse}}\ and\ \bibinfo {author} {\bibfnamefont {V.}~\bibnamefont {Elser}},\
  }\href {\doibase 10.1103/PhysRevLett.60.2531} {\bibfield  {journal} {\bibinfo
   {journal} {Phys. Rev. Lett.}\ }\textbf {\bibinfo {volume} {60}},\ \bibinfo
  {pages} {2531} (\bibinfo {year} {1988})}\BibitemShut {NoStop}%
\bibitem [{\citenamefont {Bernu}\ \emph {et~al.}(1992)\citenamefont {Bernu},
  \citenamefont {Lhuillier},\ and\ \citenamefont {Pierre}}]{Bernu1992}%
  \BibitemOpen
  \bibfield  {author} {\bibinfo {author} {\bibfnamefont {B.}~\bibnamefont
  {Bernu}}, \bibinfo {author} {\bibfnamefont {C.}~\bibnamefont {Lhuillier}}, \
  and\ \bibinfo {author} {\bibfnamefont {L.}~\bibnamefont {Pierre}},\ }\href
  {\doibase 10.1103/PhysRevLett.69.2590} {\bibfield  {journal} {\bibinfo
  {journal} {Phys. Rev. Lett.}\ }\textbf {\bibinfo {volume} {69}},\ \bibinfo
  {pages} {2590} (\bibinfo {year} {1992})}\BibitemShut {NoStop}%
\bibitem [{\citenamefont {Elstner}\ \emph {et~al.}(1993)\citenamefont
  {Elstner}, \citenamefont {Singh},\ and\ \citenamefont {Young}}]{Elstner1993}%
  \BibitemOpen
  \bibfield  {author} {\bibinfo {author} {\bibfnamefont {N.}~\bibnamefont
  {Elstner}}, \bibinfo {author} {\bibfnamefont {R.~R.~P.}\ \bibnamefont
  {Singh}}, \ and\ \bibinfo {author} {\bibfnamefont {A.~P.}\ \bibnamefont
  {Young}},\ }\href {\doibase 10.1103/PhysRevLett.71.1629} {\bibfield
  {journal} {\bibinfo  {journal} {Phys. Rev. Lett.}\ }\textbf {\bibinfo
  {volume} {71}},\ \bibinfo {pages} {1629} (\bibinfo {year}
  {1993})}\BibitemShut {NoStop}%
\bibitem [{\citenamefont {Bernu}\ \emph {et~al.}(1994)\citenamefont {Bernu},
  \citenamefont {Lecheminant}, \citenamefont {Lhuillier},\ and\ \citenamefont
  {Pierre}}]{Bernu1994}%
  \BibitemOpen
  \bibfield  {author} {\bibinfo {author} {\bibfnamefont {B.}~\bibnamefont
  {Bernu}}, \bibinfo {author} {\bibfnamefont {P.}~\bibnamefont {Lecheminant}},
  \bibinfo {author} {\bibfnamefont {C.}~\bibnamefont {Lhuillier}}, \ and\
  \bibinfo {author} {\bibfnamefont {L.}~\bibnamefont {Pierre}},\ }\href
  {\doibase 10.1103/PhysRevB.50.10048} {\bibfield  {journal} {\bibinfo
  {journal} {Phys. Rev. B}\ }\textbf {\bibinfo {volume} {50}},\ \bibinfo
  {pages} {10048} (\bibinfo {year} {1994})}\BibitemShut {NoStop}%
\bibitem [{\citenamefont {Capriotti}\ \emph {et~al.}(1999)\citenamefont
  {Capriotti}, \citenamefont {Trumper},\ and\ \citenamefont
  {Sorella}}]{Capriotti1999}%
  \BibitemOpen
  \bibfield  {author} {\bibinfo {author} {\bibfnamefont {L.}~\bibnamefont
  {Capriotti}}, \bibinfo {author} {\bibfnamefont {A.~E.}\ \bibnamefont
  {Trumper}}, \ and\ \bibinfo {author} {\bibfnamefont {S.}~\bibnamefont
  {Sorella}},\ }\href {\doibase 10.1103/PhysRevLett.82.3899} {\bibfield
  {journal} {\bibinfo  {journal} {Phys. Rev. Lett.}\ }\textbf {\bibinfo
  {volume} {82}},\ \bibinfo {pages} {3899} (\bibinfo {year}
  {1999})}\BibitemShut {NoStop}%
\bibitem [{\citenamefont {White}\ and\ \citenamefont
  {Chernyshev}(2007)}]{White2007}%
  \BibitemOpen
  \bibfield  {author} {\bibinfo {author} {\bibfnamefont {S.~R.}\ \bibnamefont
  {White}}\ and\ \bibinfo {author} {\bibfnamefont {A.~L.}\ \bibnamefont
  {Chernyshev}},\ }\href {\doibase 10.1103/PhysRevLett.99.127004} {\bibfield
  {journal} {\bibinfo  {journal} {Phys. Rev. Lett.}\ }\textbf {\bibinfo
  {volume} {99}},\ \bibinfo {pages} {127004} (\bibinfo {year}
  {2007})}\BibitemShut {NoStop}%
\bibitem [{\citenamefont {Li}\ \emph {et~al.}(2020{\natexlab{a}})\citenamefont
  {Li}, \citenamefont {Li}, \citenamefont {Zhao}, \citenamefont {Luo},\ and\
  \citenamefont {Xie}}]{Xie2020}%
  \BibitemOpen
  \bibfield  {author} {\bibinfo {author} {\bibfnamefont {Q.}~\bibnamefont
  {Li}}, \bibinfo {author} {\bibfnamefont {H.}~\bibnamefont {Li}}, \bibinfo
  {author} {\bibfnamefont {J.}~\bibnamefont {Zhao}}, \bibinfo {author}
  {\bibfnamefont {H.-G.}\ \bibnamefont {Luo}}, \ and\ \bibinfo {author}
  {\bibfnamefont {Z.~Y.}\ \bibnamefont {Xie}},\ }\href@noop {} {\  (\bibinfo
  {year} {2020}{\natexlab{a}})},\ \Eprint {http://arxiv.org/abs/2009.03765}
  {arXiv:2009.03765 [cond-mat]} \BibitemShut {NoStop}%
\bibitem [{\citenamefont {Chubukov}\ \emph {et~al.}(1994)\citenamefont
  {Chubukov}, \citenamefont {Sachdev},\ and\ \citenamefont
  {Senthil}}]{Chubukov1994}%
  \BibitemOpen
  \bibfield  {author} {\bibinfo {author} {\bibfnamefont {A.~V.}\ \bibnamefont
  {Chubukov}}, \bibinfo {author} {\bibfnamefont {S.}~\bibnamefont {Sachdev}}, \
  and\ \bibinfo {author} {\bibfnamefont {T.}~\bibnamefont {Senthil}},\ }\href
  {\doibase 10.1016/0550-3213(94)90023-X} {\bibfield  {journal} {\bibinfo
  {journal} {Nucl. Phys. B}\ }\textbf {\bibinfo {volume} {426}},\ \bibinfo
  {pages} {601} (\bibinfo {year} {1994})}\BibitemShut {NoStop}%
\bibitem [{\citenamefont {Kaneko}\ \emph {et~al.}(2014)\citenamefont {Kaneko},
  \citenamefont {Morita},\ and\ \citenamefont {Imada}}]{Kaneko14}%
  \BibitemOpen
  \bibfield  {author} {\bibinfo {author} {\bibfnamefont {R.}~\bibnamefont
  {Kaneko}}, \bibinfo {author} {\bibfnamefont {S.}~\bibnamefont {Morita}}, \
  and\ \bibinfo {author} {\bibfnamefont {M.}~\bibnamefont {Imada}},\ }\href
  {\doibase 10.7566/JPSJ.83.093707} {\bibfield  {journal} {\bibinfo  {journal}
  {J. Phys. Soc. Jpn.}\ }\textbf {\bibinfo {volume} {83}},\ \bibinfo {pages}
  {093707} (\bibinfo {year} {2014})}\BibitemShut {NoStop}%
\bibitem [{\citenamefont {Li}\ \emph {et~al.}(2015)\citenamefont {Li},
  \citenamefont {Bishop},\ and\ \citenamefont {Campbell}}]{Bishop15}%
  \BibitemOpen
  \bibfield  {author} {\bibinfo {author} {\bibfnamefont {P.~H.~Y.}\
  \bibnamefont {Li}}, \bibinfo {author} {\bibfnamefont {R.~F.}\ \bibnamefont
  {Bishop}}, \ and\ \bibinfo {author} {\bibfnamefont {C.~E.}\ \bibnamefont
  {Campbell}},\ }\href {\doibase 10.1103/PhysRevB.91.014426} {\bibfield
  {journal} {\bibinfo  {journal} {Phys. Rev. B}\ }\textbf {\bibinfo {volume}
  {91}},\ \bibinfo {pages} {014426} (\bibinfo {year} {2015})}\BibitemShut
  {NoStop}%
\bibitem [{\citenamefont {Saadatmand}\ and\ \citenamefont
  {McCulloch}(2016)}]{McCulloch2016}%
  \BibitemOpen
  \bibfield  {author} {\bibinfo {author} {\bibfnamefont {S.~N.}\ \bibnamefont
  {Saadatmand}}\ and\ \bibinfo {author} {\bibfnamefont {I.~P.}\ \bibnamefont
  {McCulloch}},\ }\href {\doibase 10.1103/PhysRevB.94.121111} {\bibfield
  {journal} {\bibinfo  {journal} {Phys. Rev. B}\ }\textbf {\bibinfo {volume}
  {94}},\ \bibinfo {pages} {121111(R)} (\bibinfo {year} {2016})}\BibitemShut
  {NoStop}%
\bibitem [{\citenamefont {Hu}\ \emph {et~al.}(2015)\citenamefont {Hu},
  \citenamefont {Gong}, \citenamefont {Zhu},\ and\ \citenamefont
  {Sheng}}]{Hu15}%
  \BibitemOpen
  \bibfield  {author} {\bibinfo {author} {\bibfnamefont {W.-J.}\ \bibnamefont
  {Hu}}, \bibinfo {author} {\bibfnamefont {S.-S.}\ \bibnamefont {Gong}},
  \bibinfo {author} {\bibfnamefont {W.}~\bibnamefont {Zhu}}, \ and\ \bibinfo
  {author} {\bibfnamefont {D.~N.}\ \bibnamefont {Sheng}},\ }\href {\doibase
  10.1103/PhysRevB.92.140403} {\bibfield  {journal} {\bibinfo  {journal} {Phys.
  Rev. B}\ }\textbf {\bibinfo {volume} {92}},\ \bibinfo {pages} {140403(R)}
  (\bibinfo {year} {2015})}\BibitemShut {NoStop}%
\bibitem [{\citenamefont {Zhu}\ and\ \citenamefont {White}(2015)}]{Zhu15}%
  \BibitemOpen
  \bibfield  {author} {\bibinfo {author} {\bibfnamefont {Z.}~\bibnamefont
  {Zhu}}\ and\ \bibinfo {author} {\bibfnamefont {S.~R.}\ \bibnamefont
  {White}},\ }\href {\doibase 10.1103/PhysRevB.92.041105} {\bibfield  {journal}
  {\bibinfo  {journal} {Phys. Rev. B}\ }\textbf {\bibinfo {volume} {92}},\
  \bibinfo {pages} {041105(R)} (\bibinfo {year} {2015})}\BibitemShut {NoStop}%
\bibitem [{\citenamefont {Iqbal}\ \emph {et~al.}(2016)\citenamefont {Iqbal},
  \citenamefont {Hu}, \citenamefont {Thomale}, \citenamefont {Poilblanc},\ and\
  \citenamefont {Becca}}]{Iqbal16}%
  \BibitemOpen
  \bibfield  {author} {\bibinfo {author} {\bibfnamefont {Y.}~\bibnamefont
  {Iqbal}}, \bibinfo {author} {\bibfnamefont {W.-J.}\ \bibnamefont {Hu}},
  \bibinfo {author} {\bibfnamefont {R.}~\bibnamefont {Thomale}}, \bibinfo
  {author} {\bibfnamefont {D.}~\bibnamefont {Poilblanc}}, \ and\ \bibinfo
  {author} {\bibfnamefont {F.}~\bibnamefont {Becca}},\ }\href {\doibase
  10.1103/PhysRevB.93.144411} {\bibfield  {journal} {\bibinfo  {journal} {Phys.
  Rev. B}\ }\textbf {\bibinfo {volume} {93}},\ \bibinfo {pages} {144411}
  (\bibinfo {year} {2016})}\BibitemShut {NoStop}%
\bibitem [{\citenamefont {Oitmaa}(2020)}]{Oitmaa2020}%
  \BibitemOpen
  \bibfield  {author} {\bibinfo {author} {\bibfnamefont {J.}~\bibnamefont
  {Oitmaa}},\ }\href {\doibase 10.1103/PhysRevB.101.214422} {\bibfield
  {journal} {\bibinfo  {journal} {Phys. Rev. B}\ }\textbf {\bibinfo {volume}
  {101}},\ \bibinfo {pages} {214422} (\bibinfo {year} {2020})}\BibitemShut
  {NoStop}%
\bibitem [{\citenamefont {Shirata}\ \emph {et~al.}(2012)\citenamefont
  {Shirata}, \citenamefont {Tanaka}, \citenamefont {Matsuo},\ and\
  \citenamefont {Kindo}}]{Shirata2012}%
  \BibitemOpen
  \bibfield  {author} {\bibinfo {author} {\bibfnamefont {Y.}~\bibnamefont
  {Shirata}}, \bibinfo {author} {\bibfnamefont {H.}~\bibnamefont {Tanaka}},
  \bibinfo {author} {\bibfnamefont {A.}~\bibnamefont {Matsuo}}, \ and\ \bibinfo
  {author} {\bibfnamefont {K.}~\bibnamefont {Kindo}},\ }\href {\doibase
  10.1103/PhysRevLett.108.057205} {\bibfield  {journal} {\bibinfo  {journal}
  {Phys. Rev. Lett.}\ }\textbf {\bibinfo {volume} {108}},\ \bibinfo {pages}
  {057205} (\bibinfo {year} {2012})}\BibitemShut {NoStop}%
\bibitem [{\citenamefont {Susuki}\ \emph {et~al.}(2013)\citenamefont {Susuki},
  \citenamefont {Kurita}, \citenamefont {Tanaka}, \citenamefont {Nojiri},
  \citenamefont {Matsuo}, \citenamefont {Kindo},\ and\ \citenamefont
  {Tanaka}}]{Susuki2013}%
  \BibitemOpen
  \bibfield  {author} {\bibinfo {author} {\bibfnamefont {T.}~\bibnamefont
  {Susuki}}, \bibinfo {author} {\bibfnamefont {N.}~\bibnamefont {Kurita}},
  \bibinfo {author} {\bibfnamefont {T.}~\bibnamefont {Tanaka}}, \bibinfo
  {author} {\bibfnamefont {H.}~\bibnamefont {Nojiri}}, \bibinfo {author}
  {\bibfnamefont {A.}~\bibnamefont {Matsuo}}, \bibinfo {author} {\bibfnamefont
  {K.}~\bibnamefont {Kindo}}, \ and\ \bibinfo {author} {\bibfnamefont
  {H.}~\bibnamefont {Tanaka}},\ }\href {\doibase
  10.1103/PhysRevLett.110.267201} {\bibfield  {journal} {\bibinfo  {journal}
  {Phys. Rev. Lett.}\ }\textbf {\bibinfo {volume} {110}},\ \bibinfo {pages}
  {267201} (\bibinfo {year} {2013})}\BibitemShut {NoStop}%
\bibitem [{\citenamefont {Ma}\ \emph {et~al.}(2016)\citenamefont {Ma},
  \citenamefont {Kamiya}, \citenamefont {Hong}, \citenamefont {Cao},
  \citenamefont {Ehlers}, \citenamefont {Tian}, \citenamefont {Batista},
  \citenamefont {Dun}, \citenamefont {Zhou},\ and\ \citenamefont
  {Matsuda}}]{Ma16}%
  \BibitemOpen
  \bibfield  {author} {\bibinfo {author} {\bibfnamefont {J.}~\bibnamefont
  {Ma}}, \bibinfo {author} {\bibfnamefont {Y.}~\bibnamefont {Kamiya}}, \bibinfo
  {author} {\bibfnamefont {T.}~\bibnamefont {Hong}}, \bibinfo {author}
  {\bibfnamefont {H.~B.}\ \bibnamefont {Cao}}, \bibinfo {author} {\bibfnamefont
  {G.}~\bibnamefont {Ehlers}}, \bibinfo {author} {\bibfnamefont
  {W.}~\bibnamefont {Tian}}, \bibinfo {author} {\bibfnamefont {C.~D.}\
  \bibnamefont {Batista}}, \bibinfo {author} {\bibfnamefont {Z.~L.}\
  \bibnamefont {Dun}}, \bibinfo {author} {\bibfnamefont {H.~D.}\ \bibnamefont
  {Zhou}}, \ and\ \bibinfo {author} {\bibfnamefont {M.}~\bibnamefont
  {Matsuda}},\ }\href {\doibase 10.1103/PhysRevLett.116.087201} {\bibfield
  {journal} {\bibinfo  {journal} {Phys. Rev. Lett.}\ }\textbf {\bibinfo
  {volume} {116}},\ \bibinfo {pages} {087201} (\bibinfo {year}
  {2016})}\BibitemShut {NoStop}%
\bibitem [{\citenamefont {Ito}\ \emph {et~al.}(2017)\citenamefont {Ito},
  \citenamefont {Kurita}, \citenamefont {Tanaka}, \citenamefont
  {Ohira-Kawamura}, \citenamefont {Nakajima}, \citenamefont {Itoh},
  \citenamefont {Kuwahara},\ and\ \citenamefont {Kakurai}}]{Ito2017}%
  \BibitemOpen
  \bibfield  {author} {\bibinfo {author} {\bibfnamefont {S.}~\bibnamefont
  {Ito}}, \bibinfo {author} {\bibfnamefont {N.}~\bibnamefont {Kurita}},
  \bibinfo {author} {\bibfnamefont {H.}~\bibnamefont {Tanaka}}, \bibinfo
  {author} {\bibfnamefont {S.}~\bibnamefont {Ohira-Kawamura}}, \bibinfo
  {author} {\bibfnamefont {K.}~\bibnamefont {Nakajima}}, \bibinfo {author}
  {\bibfnamefont {S.}~\bibnamefont {Itoh}}, \bibinfo {author} {\bibfnamefont
  {K.}~\bibnamefont {Kuwahara}}, \ and\ \bibinfo {author} {\bibfnamefont
  {K.}~\bibnamefont {Kakurai}},\ }\href {\doibase 10.1038/s41467-017-00316-x}
  {\bibfield  {journal} {\bibinfo  {journal} {Nat. Comm.}\ }\textbf {\bibinfo
  {volume} {8}},\ \bibinfo {pages} {235} (\bibinfo {year} {2017})}\BibitemShut
  {NoStop}%
\bibitem [{\citenamefont {Kamiya}\ \emph {et~al.}(2018)\citenamefont {Kamiya},
  \citenamefont {Ge}, \citenamefont {Hong}, \citenamefont {Qiu}, \citenamefont
  {Quintero-Castro}, \citenamefont {Lu}, \citenamefont {Cao}, \citenamefont
  {Matsuda}, \citenamefont {S.~Choi}, \citenamefont {Batista}, \citenamefont
  {Mourigal}, \citenamefont {D.~Zhou},\ and\ \citenamefont {Ma}}]{Kamiya}%
  \BibitemOpen
  \bibfield  {author} {\bibinfo {author} {\bibfnamefont {Y.}~\bibnamefont
  {Kamiya}}, \bibinfo {author} {\bibfnamefont {L.}~\bibnamefont {Ge}}, \bibinfo
  {author} {\bibfnamefont {T.}~\bibnamefont {Hong}}, \bibinfo {author}
  {\bibfnamefont {Y.}~\bibnamefont {Qiu}}, \bibinfo {author} {\bibfnamefont
  {D.}~\bibnamefont {Quintero-Castro}}, \bibinfo {author} {\bibfnamefont
  {Z.}~\bibnamefont {Lu}}, \bibinfo {author} {\bibfnamefont {H.}~\bibnamefont
  {Cao}}, \bibinfo {author} {\bibfnamefont {M.}~\bibnamefont {Matsuda}},
  \bibinfo {author} {\bibfnamefont {E.}~\bibnamefont {S.~Choi}}, \bibinfo
  {author} {\bibfnamefont {C.}~\bibnamefont {Batista}}, \bibinfo {author}
  {\bibfnamefont {M.}~\bibnamefont {Mourigal}}, \bibinfo {author}
  {\bibfnamefont {H.}~\bibnamefont {D.~Zhou}}, \ and\ \bibinfo {author}
  {\bibfnamefont {J.}~\bibnamefont {Ma}},\ }\href {\doibase
  10.1038/s41467-018-04914-1} {\bibfield  {journal} {\bibinfo  {journal} {Nat.
  Comm.}\ }\textbf {\bibinfo {volume} {9}},\ \bibinfo {pages} {2666} (\bibinfo
  {year} {2018})}\BibitemShut {NoStop}%
\bibitem [{\citenamefont {Macdougal}\ \emph {et~al.}(2020)\citenamefont
  {Macdougal}, \citenamefont {Williams}, \citenamefont {Prabhakaran},
  \citenamefont {Bewley}, \citenamefont {Voneshen},\ and\ \citenamefont
  {Coldea}}]{Coldea2020}%
  \BibitemOpen
  \bibfield  {author} {\bibinfo {author} {\bibfnamefont {D.}~\bibnamefont
  {Macdougal}}, \bibinfo {author} {\bibfnamefont {S.}~\bibnamefont {Williams}},
  \bibinfo {author} {\bibfnamefont {D.}~\bibnamefont {Prabhakaran}}, \bibinfo
  {author} {\bibfnamefont {R.~I.}\ \bibnamefont {Bewley}}, \bibinfo {author}
  {\bibfnamefont {D.~J.}\ \bibnamefont {Voneshen}}, \ and\ \bibinfo {author}
  {\bibfnamefont {R.}~\bibnamefont {Coldea}},\ }\href {\doibase
  10.1103/PhysRevB.102.064421} {\bibfield  {journal} {\bibinfo  {journal}
  {Phys. Rev. B}\ }\textbf {\bibinfo {volume} {102}},\ \bibinfo {pages}
  {064421} (\bibinfo {year} {2020})}\BibitemShut {NoStop}%
\bibitem [{\citenamefont {Chernyshev}\ and\ \citenamefont
  {Zhitomirsky}(2009)}]{Chernyshev09}%
  \BibitemOpen
  \bibfield  {author} {\bibinfo {author} {\bibfnamefont {A.~L.}\ \bibnamefont
  {Chernyshev}}\ and\ \bibinfo {author} {\bibfnamefont {M.~E.}\ \bibnamefont
  {Zhitomirsky}},\ }\href {\doibase 10.1103/PhysRevB.79.144416} {\bibfield
  {journal} {\bibinfo  {journal} {Phys. Rev. B}\ }\textbf {\bibinfo {volume}
  {79}},\ \bibinfo {pages} {144416} (\bibinfo {year} {2009})}\BibitemShut
  {NoStop}%
\bibitem [{\citenamefont {Mourigal}\ \emph {et~al.}(2013)\citenamefont
  {Mourigal}, \citenamefont {Fuhrman}, \citenamefont {Chernyshev},\ and\
  \citenamefont {Zhitomirsky}}]{Mourigal13}%
  \BibitemOpen
  \bibfield  {author} {\bibinfo {author} {\bibfnamefont {M.}~\bibnamefont
  {Mourigal}}, \bibinfo {author} {\bibfnamefont {W.~T.}\ \bibnamefont
  {Fuhrman}}, \bibinfo {author} {\bibfnamefont {A.~L.}\ \bibnamefont
  {Chernyshev}}, \ and\ \bibinfo {author} {\bibfnamefont {M.~E.}\ \bibnamefont
  {Zhitomirsky}},\ }\href {\doibase 10.1103/PhysRevB.88.094407} {\bibfield
  {journal} {\bibinfo  {journal} {Phys. Rev. B}\ }\textbf {\bibinfo {volume}
  {88}},\ \bibinfo {pages} {094407} (\bibinfo {year} {2013})}\BibitemShut
  {NoStop}%
\bibitem [{\citenamefont {Ghioldi}\ \emph {et~al.}(2018)\citenamefont
  {Ghioldi}, \citenamefont {Gonzalez}, \citenamefont {Zhang}, \citenamefont
  {Kamiya}, \citenamefont {Manuel}, \citenamefont {Trumper},\ and\
  \citenamefont {Batista}}]{Ghioldi18}%
  \BibitemOpen
  \bibfield  {author} {\bibinfo {author} {\bibfnamefont {E.~A.}\ \bibnamefont
  {Ghioldi}}, \bibinfo {author} {\bibfnamefont {M.~G.}\ \bibnamefont
  {Gonzalez}}, \bibinfo {author} {\bibfnamefont {S.-S.}\ \bibnamefont {Zhang}},
  \bibinfo {author} {\bibfnamefont {Y.}~\bibnamefont {Kamiya}}, \bibinfo
  {author} {\bibfnamefont {L.~O.}\ \bibnamefont {Manuel}}, \bibinfo {author}
  {\bibfnamefont {A.~E.}\ \bibnamefont {Trumper}}, \ and\ \bibinfo {author}
  {\bibfnamefont {C.~D.}\ \bibnamefont {Batista}},\ }\href {\doibase
  10.1103/PhysRevB.98.184403} {\bibfield  {journal} {\bibinfo  {journal} {Phys.
  Rev. B}\ }\textbf {\bibinfo {volume} {98}},\ \bibinfo {pages} {184403}
  (\bibinfo {year} {2018})}\BibitemShut {NoStop}%
\bibitem [{\citenamefont {Trumper}(1999)}]{Trumper1999}%
  \BibitemOpen
  \bibfield  {author} {\bibinfo {author} {\bibfnamefont {A.~E.}\ \bibnamefont
  {Trumper}},\ }\href {\doibase 10.1103/PhysRevB.60.2987} {\bibfield  {journal}
  {\bibinfo  {journal} {Phys. Rev. B}\ }\textbf {\bibinfo {volume} {60}},\
  \bibinfo {pages} {2987} (\bibinfo {year} {1999})}\BibitemShut {NoStop}%
\bibitem [{\citenamefont {Manuel}\ and\ \citenamefont
  {Ceccatto}(1999)}]{Manuel1999}%
  \BibitemOpen
  \bibfield  {author} {\bibinfo {author} {\bibfnamefont {L.~O.}\ \bibnamefont
  {Manuel}}\ and\ \bibinfo {author} {\bibfnamefont {H.~A.}\ \bibnamefont
  {Ceccatto}},\ }\href {\doibase 10.1103/PhysRevB.60.9489} {\bibfield
  {journal} {\bibinfo  {journal} {Phys. Rev. B}\ }\textbf {\bibinfo {volume}
  {60}},\ \bibinfo {pages} {9489} (\bibinfo {year} {1999})}\BibitemShut
  {NoStop}%
\bibitem [{\citenamefont {Yunoki}\ and\ \citenamefont
  {Sorella}(2006)}]{Yunoki2006}%
  \BibitemOpen
  \bibfield  {author} {\bibinfo {author} {\bibfnamefont {S.}~\bibnamefont
  {Yunoki}}\ and\ \bibinfo {author} {\bibfnamefont {S.}~\bibnamefont
  {Sorella}},\ }\href {\doibase 10.1103/PhysRevB.74.014408} {\bibfield
  {journal} {\bibinfo  {journal} {Phys. Rev. B}\ }\textbf {\bibinfo {volume}
  {74}},\ \bibinfo {pages} {014408} (\bibinfo {year} {2006})}\BibitemShut
  {NoStop}%
\bibitem [{\citenamefont {Heidarian}\ \emph {et~al.}(2009)\citenamefont
  {Heidarian}, \citenamefont {Sorella},\ and\ \citenamefont
  {Becca}}]{Heidarian09}%
  \BibitemOpen
  \bibfield  {author} {\bibinfo {author} {\bibfnamefont {D.}~\bibnamefont
  {Heidarian}}, \bibinfo {author} {\bibfnamefont {S.}~\bibnamefont {Sorella}},
  \ and\ \bibinfo {author} {\bibfnamefont {F.}~\bibnamefont {Becca}},\ }\href
  {\doibase 10.1103/PhysRevB.80.012404} {\bibfield  {journal} {\bibinfo
  {journal} {Phys. Rev. B}\ }\textbf {\bibinfo {volume} {80}},\ \bibinfo
  {pages} {012404} (\bibinfo {year} {2009})}\BibitemShut {NoStop}%
\bibitem [{\citenamefont {Thesberg}\ and\ \citenamefont
  {S\o{}rensen}(2014)}]{Mischa2014}%
  \BibitemOpen
  \bibfield  {author} {\bibinfo {author} {\bibfnamefont {M.}~\bibnamefont
  {Thesberg}}\ and\ \bibinfo {author} {\bibfnamefont {E.~S.}\ \bibnamefont
  {S\o{}rensen}},\ }\href {\doibase 10.1103/PhysRevB.90.115117} {\bibfield
  {journal} {\bibinfo  {journal} {Phys. Rev. B}\ }\textbf {\bibinfo {volume}
  {90}},\ \bibinfo {pages} {115117} (\bibinfo {year} {2014})}\BibitemShut
  {NoStop}%
\bibitem [{\citenamefont {Ghorbani}\ \emph {et~al.}(2016)\citenamefont
  {Ghorbani}, \citenamefont {Tocchio},\ and\ \citenamefont
  {Becca}}]{Ghorbani16}%
  \BibitemOpen
  \bibfield  {author} {\bibinfo {author} {\bibfnamefont {E.}~\bibnamefont
  {Ghorbani}}, \bibinfo {author} {\bibfnamefont {L.~F.}\ \bibnamefont
  {Tocchio}}, \ and\ \bibinfo {author} {\bibfnamefont {F.}~\bibnamefont
  {Becca}},\ }\href {\doibase 10.1103/PhysRevB.93.085111} {\bibfield  {journal}
  {\bibinfo  {journal} {Phys. Rev. B}\ }\textbf {\bibinfo {volume} {93}},\
  \bibinfo {pages} {085111} (\bibinfo {year} {2016})}\BibitemShut {NoStop}%
\bibitem [{\citenamefont {Coldea}\ \emph {et~al.}(2002)\citenamefont {Coldea},
  \citenamefont {Tennant}, \citenamefont {Habicht}, \citenamefont {Smeibidl},
  \citenamefont {Wolters},\ and\ \citenamefont {Tylczynski}}]{Coldea02}%
  \BibitemOpen
  \bibfield  {author} {\bibinfo {author} {\bibfnamefont {R.}~\bibnamefont
  {Coldea}}, \bibinfo {author} {\bibfnamefont {D.~A.}\ \bibnamefont {Tennant}},
  \bibinfo {author} {\bibfnamefont {K.}~\bibnamefont {Habicht}}, \bibinfo
  {author} {\bibfnamefont {P.}~\bibnamefont {Smeibidl}}, \bibinfo {author}
  {\bibfnamefont {C.}~\bibnamefont {Wolters}}, \ and\ \bibinfo {author}
  {\bibfnamefont {Z.}~\bibnamefont {Tylczynski}},\ }\href {\doibase
  10.1103/PhysRevLett.88.137203} {\bibfield  {journal} {\bibinfo  {journal}
  {Phys. Rev. Lett.}\ }\textbf {\bibinfo {volume} {88}},\ \bibinfo {pages}
  {137203} (\bibinfo {year} {2002})}\BibitemShut {NoStop}%
\bibitem [{\citenamefont {Coldea}\ \emph {et~al.}(2003)\citenamefont {Coldea},
  \citenamefont {Tennant},\ and\ \citenamefont {Tylczynski}}]{Coldea}%
  \BibitemOpen
  \bibfield  {author} {\bibinfo {author} {\bibfnamefont {R.}~\bibnamefont
  {Coldea}}, \bibinfo {author} {\bibfnamefont {D.~A.}\ \bibnamefont {Tennant}},
  \ and\ \bibinfo {author} {\bibfnamefont {Z.}~\bibnamefont {Tylczynski}},\
  }\href {\doibase 10.1103/PhysRevB.68.134424} {\bibfield  {journal} {\bibinfo
  {journal} {Phys. Rev. B}\ }\textbf {\bibinfo {volume} {68}},\ \bibinfo
  {pages} {134424} (\bibinfo {year} {2003})}\BibitemShut {NoStop}%
\bibitem [{\citenamefont {Fj\ae{}restad}\ \emph {et~al.}(2007)\citenamefont
  {Fj\ae{}restad}, \citenamefont {Zheng}, \citenamefont {Singh}, \citenamefont
  {McKenzie},\ and\ \citenamefont {Coldea}}]{Fjaerestad07}%
  \BibitemOpen
  \bibfield  {author} {\bibinfo {author} {\bibfnamefont {J.~O.}\ \bibnamefont
  {Fj\ae{}restad}}, \bibinfo {author} {\bibfnamefont {W.}~\bibnamefont
  {Zheng}}, \bibinfo {author} {\bibfnamefont {R.~R.~P.}\ \bibnamefont {Singh}},
  \bibinfo {author} {\bibfnamefont {R.~H.}\ \bibnamefont {McKenzie}}, \ and\
  \bibinfo {author} {\bibfnamefont {R.}~\bibnamefont {Coldea}},\ }\href
  {\doibase 10.1103/PhysRevB.75.174447} {\bibfield  {journal} {\bibinfo
  {journal} {Phys. Rev. B}\ }\textbf {\bibinfo {volume} {75}},\ \bibinfo
  {pages} {174447} (\bibinfo {year} {2007})}\BibitemShut {NoStop}%
\bibitem [{\citenamefont {Starykh}\ \emph {et~al.}(2010)\citenamefont
  {Starykh}, \citenamefont {Katsura},\ and\ \citenamefont
  {Balents}}]{Starykh10}%
  \BibitemOpen
  \bibfield  {author} {\bibinfo {author} {\bibfnamefont {O.~A.}\ \bibnamefont
  {Starykh}}, \bibinfo {author} {\bibfnamefont {H.}~\bibnamefont {Katsura}}, \
  and\ \bibinfo {author} {\bibfnamefont {L.}~\bibnamefont {Balents}},\ }\href
  {\doibase 10.1103/PhysRevB.82.014421} {\bibfield  {journal} {\bibinfo
  {journal} {Phys. Rev. B}\ }\textbf {\bibinfo {volume} {82}},\ \bibinfo
  {pages} {014421} (\bibinfo {year} {2010})}\BibitemShut {NoStop}%
\bibitem [{\citenamefont {Zheng}\ \emph {et~al.}(2005)\citenamefont {Zheng},
  \citenamefont {Singh}, \citenamefont {McKenzie},\ and\ \citenamefont
  {Coldea}}]{Zheng05}%
  \BibitemOpen
  \bibfield  {author} {\bibinfo {author} {\bibfnamefont {W.}~\bibnamefont
  {Zheng}}, \bibinfo {author} {\bibfnamefont {R.~R.~P.}\ \bibnamefont {Singh}},
  \bibinfo {author} {\bibfnamefont {R.~H.}\ \bibnamefont {McKenzie}}, \ and\
  \bibinfo {author} {\bibfnamefont {R.}~\bibnamefont {Coldea}},\ }\href
  {\doibase 10.1103/PhysRevB.71.134422} {\bibfield  {journal} {\bibinfo
  {journal} {Phys. Rev. B}\ }\textbf {\bibinfo {volume} {71}},\ \bibinfo
  {pages} {134422} (\bibinfo {year} {2005})}\BibitemShut {NoStop}%
\bibitem [{\citenamefont {Alicea}\ \emph {et~al.}(2005)\citenamefont {Alicea},
  \citenamefont {Motrunich},\ and\ \citenamefont {Fisher}}]{Alicea2005}%
  \BibitemOpen
  \bibfield  {author} {\bibinfo {author} {\bibfnamefont {J.}~\bibnamefont
  {Alicea}}, \bibinfo {author} {\bibfnamefont {O.~I.}\ \bibnamefont
  {Motrunich}}, \ and\ \bibinfo {author} {\bibfnamefont {M.~P.~A.}\
  \bibnamefont {Fisher}},\ }\href {\doibase 10.1103/PhysRevLett.95.247203}
  {\bibfield  {journal} {\bibinfo  {journal} {Phys. Rev. Lett.}\ }\textbf
  {\bibinfo {volume} {95}},\ \bibinfo {pages} {247203} (\bibinfo {year}
  {2005})}\BibitemShut {NoStop}%
\bibitem [{\citenamefont {Zheng}\ \emph {et~al.}(2006)\citenamefont {Zheng},
  \citenamefont {Fj\ae{}restad}, \citenamefont {Singh}, \citenamefont
  {McKenzie},\ and\ \citenamefont {Coldea}}]{Zheng}%
  \BibitemOpen
  \bibfield  {author} {\bibinfo {author} {\bibfnamefont {W.}~\bibnamefont
  {Zheng}}, \bibinfo {author} {\bibfnamefont {J.~O.}\ \bibnamefont
  {Fj\ae{}restad}}, \bibinfo {author} {\bibfnamefont {R.~R.~P.}\ \bibnamefont
  {Singh}}, \bibinfo {author} {\bibfnamefont {R.~H.}\ \bibnamefont {McKenzie}},
  \ and\ \bibinfo {author} {\bibfnamefont {R.}~\bibnamefont {Coldea}},\ }\href
  {\doibase 10.1103/PhysRevLett.96.057201} {\bibfield  {journal} {\bibinfo
  {journal} {Phys. Rev. Lett.}\ }\textbf {\bibinfo {volume} {96}},\ \bibinfo
  {pages} {057201} (\bibinfo {year} {2006})}\BibitemShut {NoStop}%
\bibitem [{\citenamefont {Kohno}\ \emph {et~al.}(2007)\citenamefont {Kohno},
  \citenamefont {Starykh},\ and\ \citenamefont {Balents}}]{Kohno}%
  \BibitemOpen
  \bibfield  {author} {\bibinfo {author} {\bibfnamefont {M.}~\bibnamefont
  {Kohno}}, \bibinfo {author} {\bibfnamefont {O.~A.}\ \bibnamefont {Starykh}},
  \ and\ \bibinfo {author} {\bibfnamefont {L.}~\bibnamefont {Balents}},\ }\href
  {\doibase 10.1038/nphys749} {\bibfield  {journal} {\bibinfo  {journal} {Nat.
  Phys.}\ }\textbf {\bibinfo {volume} {3}},\ \bibinfo {pages} {790} (\bibinfo
  {year} {2007})}\BibitemShut {NoStop}%
\bibitem [{\citenamefont {Gonzalez}\ \emph {et~al.}(2017)\citenamefont
  {Gonzalez}, \citenamefont {Ghioldi}, \citenamefont {Gazza}, \citenamefont
  {Manuel},\ and\ \citenamefont {Trumper}}]{Gonzalez17}%
  \BibitemOpen
  \bibfield  {author} {\bibinfo {author} {\bibfnamefont {M.~G.}\ \bibnamefont
  {Gonzalez}}, \bibinfo {author} {\bibfnamefont {E.~A.}\ \bibnamefont
  {Ghioldi}}, \bibinfo {author} {\bibfnamefont {C.~J.}\ \bibnamefont {Gazza}},
  \bibinfo {author} {\bibfnamefont {L.~O.}\ \bibnamefont {Manuel}}, \ and\
  \bibinfo {author} {\bibfnamefont {A.~E.}\ \bibnamefont {Trumper}},\ }\href
  {\doibase 10.1103/PhysRevB.96.174423} {\bibfield  {journal} {\bibinfo
  {journal} {Phys. Rev. B}\ }\textbf {\bibinfo {volume} {96}},\ \bibinfo
  {pages} {174423} (\bibinfo {year} {2017})}\BibitemShut {NoStop}%
\bibitem [{\citenamefont {Hembacher}\ \emph {et~al.}(2018)\citenamefont
  {Hembacher}, \citenamefont {Badrtdinov}, \citenamefont {Ding}, \citenamefont
  {Sobczak}, \citenamefont {Ritter}, \citenamefont {Mazurenko},\ and\
  \citenamefont {Tsirlin}}]{Hembacher18}%
  \BibitemOpen
  \bibfield  {author} {\bibinfo {author} {\bibfnamefont {J.}~\bibnamefont
  {Hembacher}}, \bibinfo {author} {\bibfnamefont {D.~I.}\ \bibnamefont
  {Badrtdinov}}, \bibinfo {author} {\bibfnamefont {L.}~\bibnamefont {Ding}},
  \bibinfo {author} {\bibfnamefont {Z.}~\bibnamefont {Sobczak}}, \bibinfo
  {author} {\bibfnamefont {C.}~\bibnamefont {Ritter}}, \bibinfo {author}
  {\bibfnamefont {V.~V.}\ \bibnamefont {Mazurenko}}, \ and\ \bibinfo {author}
  {\bibfnamefont {A.~A.}\ \bibnamefont {Tsirlin}},\ }\href {\doibase
  10.1103/PhysRevB.98.094406} {\bibfield  {journal} {\bibinfo  {journal} {Phys.
  Rev. B}\ }\textbf {\bibinfo {volume} {98}},\ \bibinfo {pages} {094406}
  (\bibinfo {year} {2018})}\BibitemShut {NoStop}%
\bibitem [{\citenamefont {Abdeldaim}\ \emph {et~al.}(2019)\citenamefont
  {Abdeldaim}, \citenamefont {Badrtdinov}, \citenamefont {Gibbs}, \citenamefont
  {Manuel}, \citenamefont {Walker}, \citenamefont {Le}, \citenamefont {Wu},
  \citenamefont {Wardecki}, \citenamefont {Eriksson}, \citenamefont {Kvashnin},
  \citenamefont {Tsirlin},\ and\ \citenamefont {Nilsen}}]{Abdeldaim19}%
  \BibitemOpen
  \bibfield  {author} {\bibinfo {author} {\bibfnamefont {A.~H.}\ \bibnamefont
  {Abdeldaim}}, \bibinfo {author} {\bibfnamefont {D.~I.}\ \bibnamefont
  {Badrtdinov}}, \bibinfo {author} {\bibfnamefont {A.~S.}\ \bibnamefont
  {Gibbs}}, \bibinfo {author} {\bibfnamefont {P.}~\bibnamefont {Manuel}},
  \bibinfo {author} {\bibfnamefont {H.~C.}\ \bibnamefont {Walker}}, \bibinfo
  {author} {\bibfnamefont {M.~D.}\ \bibnamefont {Le}}, \bibinfo {author}
  {\bibfnamefont {C.~H.}\ \bibnamefont {Wu}}, \bibinfo {author} {\bibfnamefont
  {D.}~\bibnamefont {Wardecki}}, \bibinfo {author} {\bibfnamefont {S.-G.}\
  \bibnamefont {Eriksson}}, \bibinfo {author} {\bibfnamefont {Y.~O.}\
  \bibnamefont {Kvashnin}}, \bibinfo {author} {\bibfnamefont {A.~A.}\
  \bibnamefont {Tsirlin}}, \ and\ \bibinfo {author} {\bibfnamefont {G.~J.}\
  \bibnamefont {Nilsen}},\ }\href {\doibase 10.1103/PhysRevB.100.214427}
  {\bibfield  {journal} {\bibinfo  {journal} {Phys. Rev. B}\ }\textbf {\bibinfo
  {volume} {100}},\ \bibinfo {pages} {214427} (\bibinfo {year}
  {2019})}\BibitemShut {NoStop}%
\bibitem [{\citenamefont {Arovas}\ and\ \citenamefont
  {Auerbach}(1988)}]{Arovas88}%
  \BibitemOpen
  \bibfield  {author} {\bibinfo {author} {\bibfnamefont {D.~P.}\ \bibnamefont
  {Arovas}}\ and\ \bibinfo {author} {\bibfnamefont {A.}~\bibnamefont
  {Auerbach}},\ }\href {\doibase 10.1103/PhysRevB.38.316} {\bibfield  {journal}
  {\bibinfo  {journal} {Phys. Rev. B}\ }\textbf {\bibinfo {volume} {38}},\
  \bibinfo {pages} {316} (\bibinfo {year} {1988})}\BibitemShut {NoStop}%
\bibitem [{\citenamefont {Auerbach}(1994)}]{Auerbach1994}%
  \BibitemOpen
  \bibfield  {author} {\bibinfo {author} {\bibfnamefont {A.}~\bibnamefont
  {Auerbach}},\ }\href@noop {} {\emph {\bibinfo {title} {Interacting electrons
  and quantum magnetism}}}\ (\bibinfo  {publisher} {Springer-Verlag},\ \bibinfo
  {address} {New York},\ \bibinfo {year} {1994})\BibitemShut {NoStop}%
\bibitem [{\citenamefont {Trumper}\ \emph {et~al.}(1997)\citenamefont
  {Trumper}, \citenamefont {Manuel}, \citenamefont {Gazza},\ and\ \citenamefont
  {Ceccatto}}]{Trumper97}%
  \BibitemOpen
  \bibfield  {author} {\bibinfo {author} {\bibfnamefont {A.~E.}\ \bibnamefont
  {Trumper}}, \bibinfo {author} {\bibfnamefont {L.~O.}\ \bibnamefont {Manuel}},
  \bibinfo {author} {\bibfnamefont {C.~J.}\ \bibnamefont {Gazza}}, \ and\
  \bibinfo {author} {\bibfnamefont {H.~A.}\ \bibnamefont {Ceccatto}},\ }\href
  {\doibase 10.1103/PhysRevLett.78.2216} {\bibfield  {journal} {\bibinfo
  {journal} {Phys. Rev. Lett.}\ }\textbf {\bibinfo {volume} {78}},\ \bibinfo
  {pages} {2216} (\bibinfo {year} {1997})}\BibitemShut {NoStop}%
\bibitem [{\citenamefont {Read}\ and\ \citenamefont
  {Sachdev}(1991)}]{Read1991}%
  \BibitemOpen
  \bibfield  {author} {\bibinfo {author} {\bibfnamefont {N.}~\bibnamefont
  {Read}}\ and\ \bibinfo {author} {\bibfnamefont {S.}~\bibnamefont {Sachdev}},\
  }\href {\doibase 10.1103/PhysRevLett.66.1773} {\bibfield  {journal} {\bibinfo
   {journal} {Phys. Rev. Lett.}\ }\textbf {\bibinfo {volume} {66}},\ \bibinfo
  {pages} {1773} (\bibinfo {year} {1991})}\BibitemShut {NoStop}%
\bibitem [{\citenamefont {Merino}\ \emph {et~al.}(2014)\citenamefont {Merino},
  \citenamefont {Holt},\ and\ \citenamefont {Powell}}]{Merino14}%
  \BibitemOpen
  \bibfield  {author} {\bibinfo {author} {\bibfnamefont {J.}~\bibnamefont
  {Merino}}, \bibinfo {author} {\bibfnamefont {M.}~\bibnamefont {Holt}}, \ and\
  \bibinfo {author} {\bibfnamefont {B.~J.}\ \bibnamefont {Powell}},\ }\href
  {\doibase 10.1103/PhysRevB.89.245112} {\bibfield  {journal} {\bibinfo
  {journal} {Phys. Rev. B}\ }\textbf {\bibinfo {volume} {89}},\ \bibinfo
  {pages} {245112} (\bibinfo {year} {2014})}\BibitemShut {NoStop}%
\bibitem [{\citenamefont {Ceccatto}\ \emph {et~al.}(1993)\citenamefont
  {Ceccatto}, \citenamefont {Gazza},\ and\ \citenamefont
  {Trumper}}]{Ceccatto93}%
  \BibitemOpen
  \bibfield  {author} {\bibinfo {author} {\bibfnamefont {H.~A.}\ \bibnamefont
  {Ceccatto}}, \bibinfo {author} {\bibfnamefont {C.~J.}\ \bibnamefont {Gazza}},
  \ and\ \bibinfo {author} {\bibfnamefont {A.~E.}\ \bibnamefont {Trumper}},\
  }\href {\doibase 10.1103/PhysRevB.47.12329} {\bibfield  {journal} {\bibinfo
  {journal} {Phys. Rev. B}\ }\textbf {\bibinfo {volume} {47}},\ \bibinfo
  {pages} {12329} (\bibinfo {year} {1993})}\BibitemShut {NoStop}%
\bibitem [{\citenamefont {Wang}\ and\ \citenamefont
  {Vishwanath}(2006)}]{Wang2006}%
  \BibitemOpen
  \bibfield  {author} {\bibinfo {author} {\bibfnamefont {F.}~\bibnamefont
  {Wang}}\ and\ \bibinfo {author} {\bibfnamefont {A.}~\bibnamefont
  {Vishwanath}},\ }\href {\doibase 10.1103/PhysRevB.74.174423} {\bibfield
  {journal} {\bibinfo  {journal} {Phys. Rev. B}\ }\textbf {\bibinfo {volume}
  {74}},\ \bibinfo {pages} {174423} (\bibinfo {year} {2006})}\BibitemShut
  {NoStop}%
\bibitem [{\citenamefont {Bauer}\ and\ \citenamefont
  {Fj\ae{}restad}(2017)}]{Bauer17}%
  \BibitemOpen
  \bibfield  {author} {\bibinfo {author} {\bibfnamefont {D.-V.}\ \bibnamefont
  {Bauer}}\ and\ \bibinfo {author} {\bibfnamefont {J.~O.}\ \bibnamefont
  {Fj\ae{}restad}},\ }\href {\doibase 10.1103/PhysRevB.96.165141} {\bibfield
  {journal} {\bibinfo  {journal} {Phys. Rev. B}\ }\textbf {\bibinfo {volume}
  {96}},\ \bibinfo {pages} {165141} (\bibinfo {year} {2017})}\BibitemShut
  {NoStop}%
\bibitem [{\citenamefont {Flint}\ and\ \citenamefont
  {Coleman}(2009)}]{Flint09}%
  \BibitemOpen
  \bibfield  {author} {\bibinfo {author} {\bibfnamefont {R.}~\bibnamefont
  {Flint}}\ and\ \bibinfo {author} {\bibfnamefont {P.}~\bibnamefont
  {Coleman}},\ }\href {\doibase 10.1103/PhysRevB.79.014424} {\bibfield
  {journal} {\bibinfo  {journal} {Phys. Rev. B}\ }\textbf {\bibinfo {volume}
  {79}},\ \bibinfo {pages} {014424} (\bibinfo {year} {2009})}\BibitemShut
  {NoStop}%
\bibitem [{\citenamefont {Mezio}\ \emph {et~al.}(2011)\citenamefont {Mezio},
  \citenamefont {Sposetti}, \citenamefont {Manuel},\ and\ \citenamefont
  {Trumper}}]{Mezio11}%
  \BibitemOpen
  \bibfield  {author} {\bibinfo {author} {\bibfnamefont {A.}~\bibnamefont
  {Mezio}}, \bibinfo {author} {\bibfnamefont {C.~N.}\ \bibnamefont {Sposetti}},
  \bibinfo {author} {\bibfnamefont {L.~O.}\ \bibnamefont {Manuel}}, \ and\
  \bibinfo {author} {\bibfnamefont {A.~E.}\ \bibnamefont {Trumper}},\ }\href
  {\doibase 10.1209/0295-5075/94/47001} {\bibfield  {journal} {\bibinfo
  {journal} {EPL}\ }\textbf {\bibinfo {volume} {94}},\ \bibinfo {pages} {47001}
  (\bibinfo {year} {2011})}\BibitemShut {NoStop}%
\bibitem [{\citenamefont {Zhang}\ \emph {et~al.}(2019)\citenamefont {Zhang},
  \citenamefont {Ghioldi}, \citenamefont {Kamiya}, \citenamefont {Manuel},
  \citenamefont {Trumper},\ and\ \citenamefont {Batista}}]{Zhang}%
  \BibitemOpen
  \bibfield  {author} {\bibinfo {author} {\bibfnamefont {S.-S.}\ \bibnamefont
  {Zhang}}, \bibinfo {author} {\bibfnamefont {E.~A.}\ \bibnamefont {Ghioldi}},
  \bibinfo {author} {\bibfnamefont {Y.}~\bibnamefont {Kamiya}}, \bibinfo
  {author} {\bibfnamefont {L.~O.}\ \bibnamefont {Manuel}}, \bibinfo {author}
  {\bibfnamefont {A.~E.}\ \bibnamefont {Trumper}}, \ and\ \bibinfo {author}
  {\bibfnamefont {C.~D.}\ \bibnamefont {Batista}},\ }\href {\doibase
  10.1103/PhysRevB.100.104431} {\bibfield  {journal} {\bibinfo  {journal}
  {Phys. Rev. B}\ }\textbf {\bibinfo {volume} {100}},\ \bibinfo {pages}
  {104431} (\bibinfo {year} {2019})}\BibitemShut {NoStop}%
\bibitem [{\citenamefont {Hirsch}\ and\ \citenamefont {Tang}(1989)}]{Hirsch89}%
  \BibitemOpen
  \bibfield  {author} {\bibinfo {author} {\bibfnamefont {J.~E.}\ \bibnamefont
  {Hirsch}}\ and\ \bibinfo {author} {\bibfnamefont {S.}~\bibnamefont {Tang}},\
  }\href {\doibase 10.1103/PhysRevB.39.2850} {\bibfield  {journal} {\bibinfo
  {journal} {Phys. Rev. B}\ }\textbf {\bibinfo {volume} {39}},\ \bibinfo
  {pages} {2850} (\bibinfo {year} {1989})}\BibitemShut {NoStop}%
\bibitem [{\citenamefont {Sarker}\ \emph {et~al.}(1989)\citenamefont {Sarker},
  \citenamefont {Jayaprakash}, \citenamefont {Krishnamurthy},\ and\
  \citenamefont {Ma}}]{Sarker}%
  \BibitemOpen
  \bibfield  {author} {\bibinfo {author} {\bibfnamefont {S.}~\bibnamefont
  {Sarker}}, \bibinfo {author} {\bibfnamefont {C.}~\bibnamefont {Jayaprakash}},
  \bibinfo {author} {\bibfnamefont {H.~R.}\ \bibnamefont {Krishnamurthy}}, \
  and\ \bibinfo {author} {\bibfnamefont {M.}~\bibnamefont {Ma}},\ }\href
  {\doibase 10.1103/PhysRevB.40.5028} {\bibfield  {journal} {\bibinfo
  {journal} {Phys. Rev. B}\ }\textbf {\bibinfo {volume} {40}},\ \bibinfo
  {pages} {5028} (\bibinfo {year} {1989})}\BibitemShut {NoStop}%
\bibitem [{\citenamefont {Chandra}\ \emph {et~al.}(1990)\citenamefont
  {Chandra}, \citenamefont {Coleman},\ and\ \citenamefont
  {Larkin}}]{Chandra90}%
  \BibitemOpen
  \bibfield  {author} {\bibinfo {author} {\bibfnamefont {P.}~\bibnamefont
  {Chandra}}, \bibinfo {author} {\bibfnamefont {P.}~\bibnamefont {Coleman}}, \
  and\ \bibinfo {author} {\bibfnamefont {A.~I.}\ \bibnamefont {Larkin}},\
  }\href {\doibase 10.1088/0953-8984/2/39/008} {\bibfield  {journal} {\bibinfo
  {journal} {J. Phys. Condens. Matter}\ }\textbf {\bibinfo {volume} {2}},\
  \bibinfo {pages} {7933} (\bibinfo {year} {1990})}\BibitemShut {NoStop}%
\bibitem [{\citenamefont {Manuel}\ \emph {et~al.}(1998)\citenamefont {Manuel},
  \citenamefont {Trumper},\ and\ \citenamefont {Ceccatto}}]{Manueltriang}%
  \BibitemOpen
  \bibfield  {author} {\bibinfo {author} {\bibfnamefont {L.~O.}\ \bibnamefont
  {Manuel}}, \bibinfo {author} {\bibfnamefont {A.~E.}\ \bibnamefont {Trumper}},
  \ and\ \bibinfo {author} {\bibfnamefont {H.~A.}\ \bibnamefont {Ceccatto}},\
  }\href {\doibase 10.1103/PhysRevB.57.8348} {\bibfield  {journal} {\bibinfo
  {journal} {Phys. Rev. B}\ }\textbf {\bibinfo {volume} {57}},\ \bibinfo
  {pages} {8348} (\bibinfo {year} {1998})}\BibitemShut {NoStop}%
\bibitem [{\citenamefont {Wen}(2002)}]{Wen2002}%
  \BibitemOpen
  \bibfield  {author} {\bibinfo {author} {\bibfnamefont {X.-G.}\ \bibnamefont
  {Wen}},\ }\href {\doibase 10.1103/PhysRevB.65.165113} {\bibfield  {journal}
  {\bibinfo  {journal} {Phys. Rev. B}\ }\textbf {\bibinfo {volume} {65}},\
  \bibinfo {pages} {165113} (\bibinfo {year} {2002})}\BibitemShut {NoStop}%
\bibitem [{\citenamefont {Messio}\ \emph {et~al.}(2013)\citenamefont {Messio},
  \citenamefont {Lhuillier},\ and\ \citenamefont {Misguich}}]{Messio}%
  \BibitemOpen
  \bibfield  {author} {\bibinfo {author} {\bibfnamefont {L.}~\bibnamefont
  {Messio}}, \bibinfo {author} {\bibfnamefont {C.}~\bibnamefont {Lhuillier}}, \
  and\ \bibinfo {author} {\bibfnamefont {G.}~\bibnamefont {Misguich}},\ }\href
  {\doibase 10.1103/PhysRevB.87.125127} {\bibfield  {journal} {\bibinfo
  {journal} {Phys. Rev. B}\ }\textbf {\bibinfo {volume} {87}},\ \bibinfo
  {pages} {125127} (\bibinfo {year} {2013})}\BibitemShut {NoStop}%
\bibitem [{\citenamefont {Lu}(2016)}]{Lu16}%
  \BibitemOpen
  \bibfield  {author} {\bibinfo {author} {\bibfnamefont {Y.-M.}\ \bibnamefont
  {Lu}},\ }\href {\doibase 10.1103/PhysRevB.93.165113} {\bibfield  {journal}
  {\bibinfo  {journal} {Phys. Rev. B}\ }\textbf {\bibinfo {volume} {93}},\
  \bibinfo {pages} {165113} (\bibinfo {year} {2016})}\BibitemShut {NoStop}%
\bibitem [{\citenamefont {Li}\ \emph {et~al.}(2012)\citenamefont {Li},
  \citenamefont {Becca}, \citenamefont {Hu},\ and\ \citenamefont
  {Sorella}}]{Sorella2012}%
  \BibitemOpen
  \bibfield  {author} {\bibinfo {author} {\bibfnamefont {T.}~\bibnamefont
  {Li}}, \bibinfo {author} {\bibfnamefont {F.}~\bibnamefont {Becca}}, \bibinfo
  {author} {\bibfnamefont {W.}~\bibnamefont {Hu}}, \ and\ \bibinfo {author}
  {\bibfnamefont {S.}~\bibnamefont {Sorella}},\ }\href {\doibase
  10.1103/PhysRevB.86.075111} {\bibfield  {journal} {\bibinfo  {journal} {Phys.
  Rev. B}\ }\textbf {\bibinfo {volume} {86}},\ \bibinfo {pages} {075111}
  (\bibinfo {year} {2012})}\BibitemShut {NoStop}%
\bibitem [{\citenamefont {Zhu}\ \emph {et~al.}(2018)\citenamefont {Zhu},
  \citenamefont {Maksimov}, \citenamefont {White},\ and\ \citenamefont
  {Chernyshev}}]{White2018}%
  \BibitemOpen
  \bibfield  {author} {\bibinfo {author} {\bibfnamefont {Z.}~\bibnamefont
  {Zhu}}, \bibinfo {author} {\bibfnamefont {P.~A.}\ \bibnamefont {Maksimov}},
  \bibinfo {author} {\bibfnamefont {S.~R.}\ \bibnamefont {White}}, \ and\
  \bibinfo {author} {\bibfnamefont {A.~L.}\ \bibnamefont {Chernyshev}},\ }\href
  {\doibase 10.1103/PhysRevLett.120.207203} {\bibfield  {journal} {\bibinfo
  {journal} {Phys. Rev. Lett.}\ }\textbf {\bibinfo {volume} {120}},\ \bibinfo
  {pages} {207203} (\bibinfo {year} {2018})}\BibitemShut {NoStop}%
\bibitem [{\citenamefont {Li}\ \emph {et~al.}(2020{\natexlab{b}})\citenamefont
  {Li}, \citenamefont {Gegenwart},\ and\ \citenamefont {Tsirlin}}]{li2020}%
  \BibitemOpen
  \bibfield  {author} {\bibinfo {author} {\bibfnamefont {Y.}~\bibnamefont
  {Li}}, \bibinfo {author} {\bibfnamefont {P.}~\bibnamefont {Gegenwart}}, \
  and\ \bibinfo {author} {\bibfnamefont {A.~A.}\ \bibnamefont {Tsirlin}},\
  }\href {\doibase 10.1088/1361-648x/ab724e} {\bibfield  {journal} {\bibinfo
  {journal} {J. Phys. Condens. Matter}\ }\textbf {\bibinfo {volume} {32}},\
  \bibinfo {pages} {224004} (\bibinfo {year} {2020}{\natexlab{b}})}\BibitemShut
  {NoStop}%
\bibitem [{\citenamefont {Ferrari}\ and\ \citenamefont {Becca}(2019)}]{Becca}%
  \BibitemOpen
  \bibfield  {author} {\bibinfo {author} {\bibfnamefont {F.}~\bibnamefont
  {Ferrari}}\ and\ \bibinfo {author} {\bibfnamefont {F.}~\bibnamefont
  {Becca}},\ }\href {\doibase 10.1103/PhysRevX.9.031026} {\bibfield  {journal}
  {\bibinfo  {journal} {Phys. Rev. X}\ }\textbf {\bibinfo {volume} {9}},\
  \bibinfo {pages} {031026} (\bibinfo {year} {2019})}\BibitemShut {NoStop}%
\bibitem [{\citenamefont {Wu}\ \emph {et~al.}(2020)\citenamefont {Wu},
  \citenamefont {Yao},\ and\ \citenamefont {Wu}}]{Wu2020}%
  \BibitemOpen
  \bibfield  {author} {\bibinfo {author} {\bibfnamefont {M.}~\bibnamefont
  {Wu}}, \bibinfo {author} {\bibfnamefont {D.-X.}\ \bibnamefont {Yao}}, \ and\
  \bibinfo {author} {\bibfnamefont {H.-Q.}\ \bibnamefont {Wu}},\ }\href@noop {}
  {\  (\bibinfo {year} {2020})},\ \Eprint {http://arxiv.org/abs/2008.08751}
  {arXiv:2008.08751 [cond-mat]} \BibitemShut {NoStop}%
\end{thebibliography}%

\end{document}